\documentclass[sigconf]{acmart}

\AtBeginDocument{%
  }

\setcopyright{acmlicensed}
\copyrightyear{2026}
\acmYear{2026}
\setcopyright{cc}
\setcctype{by}
\acmConference[CHI '26]{Proceedings of the 2026 CHI Conference on Human Factors in Computing Systems}{April 13--17, 2026}{Barcelona, Spain}
\acmBooktitle{Proceedings of the 2026 CHI Conference on Human Factors in Computing Systems (CHI '26), April 13--17, 2026, Barcelona, Spain}
\acmDOI{10.1145/3772318.3791121}
\acmISBN{979-8-4007-2278-3/2026/04}

\usepackage{enumitem}

\usepackage{subcaption}
\usepackage{multirow}
\usepackage{cleveref}
\usepackage{soul}

\definecolor{headerblue}{RGB}{220,230,241}

\newcommand{\xinyi}[1]{#1}

\newcommand{\highlight}[1]{#1}

\renewcommand\arraystretch{1.2}
\begin{document}

\title{AI-Mediated Feedback Improves Student Revisions: A Randomized Trial with FeedbackWriter in a Large Undergraduate Course}

\author{Xinyi Lu}
\affiliation{
 \institution{University of Michigan}
 \city{Ann Arbor}
 \state{Michigan}
  \country{USA}
}
\email{lwlxy@umich.edu}

\author{Kexin Phyllis Ju}
\affiliation{
 \institution{University of Michigan}
 \city{Ann Arbor}
 \state{Michigan}
  \country{USA}
}
\email{kexinju@umich.edu}

\author{Mitchell Dudley}
\affiliation{
 \institution{University of Michigan}
 \city{Ann Arbor}
 \state{Michigan}
  \country{USA}
}
\email{mrdudley@umich.edu}

\author{Larissa Sano}
\affiliation{
 \institution{University of Michigan}
 \city{Ann Arbor}
 \state{Michigan}
  \country{USA}
}
\email{llubomud@umich.edu}

\author{Xu Wang}
\affiliation{
 \institution{University of Michigan}
 \city{Ann Arbor}
 \state{Michigan}
 \country{USA}
}
\email{xwanghci@umich.edu}




\begin{abstract}
Despite growing interest in using LLMs to generate feedback on students’ writing, little is known about how students respond to AI-mediated versus human-provided feedback. We address this gap through a randomized controlled trial in a large introductory economics course (N=354), where we introduce and deploy FeedbackWriter---a system that generates AI suggestions to teaching assistants (TAs) while they provide feedback on students’ knowledge-intensive essays. TAs have the full capacity to adopt, edit, or dismiss the suggestions. Students were randomly assigned to receive either handwritten feedback from TAs (baseline) or AI-mediated feedback where TAs received suggestions from FeedbackWriter. Students revise their drafts based on the feedback, which is further graded. In total, 1,366 essays were graded using the system. We found that students receiving AI-mediated feedback produced significantly higher-quality revisions, with gains increasing as TAs adopted more AI suggestions. TAs found the AI suggestions useful for spotting gaps and clarifying rubrics.

\end{abstract}

\ccsdesc[500]{Applied computing~Computer-assisted instruction}
\ccsdesc[500]{Human-centered computing~Empirical studies in HCI}

\keywords{Randomized Controlled Trial, AI-mediated Feedback, Feedback Provision, Human-AI Interaction}

\begin{teaserfigure}
  \centering
  \includegraphics[width=\textwidth]{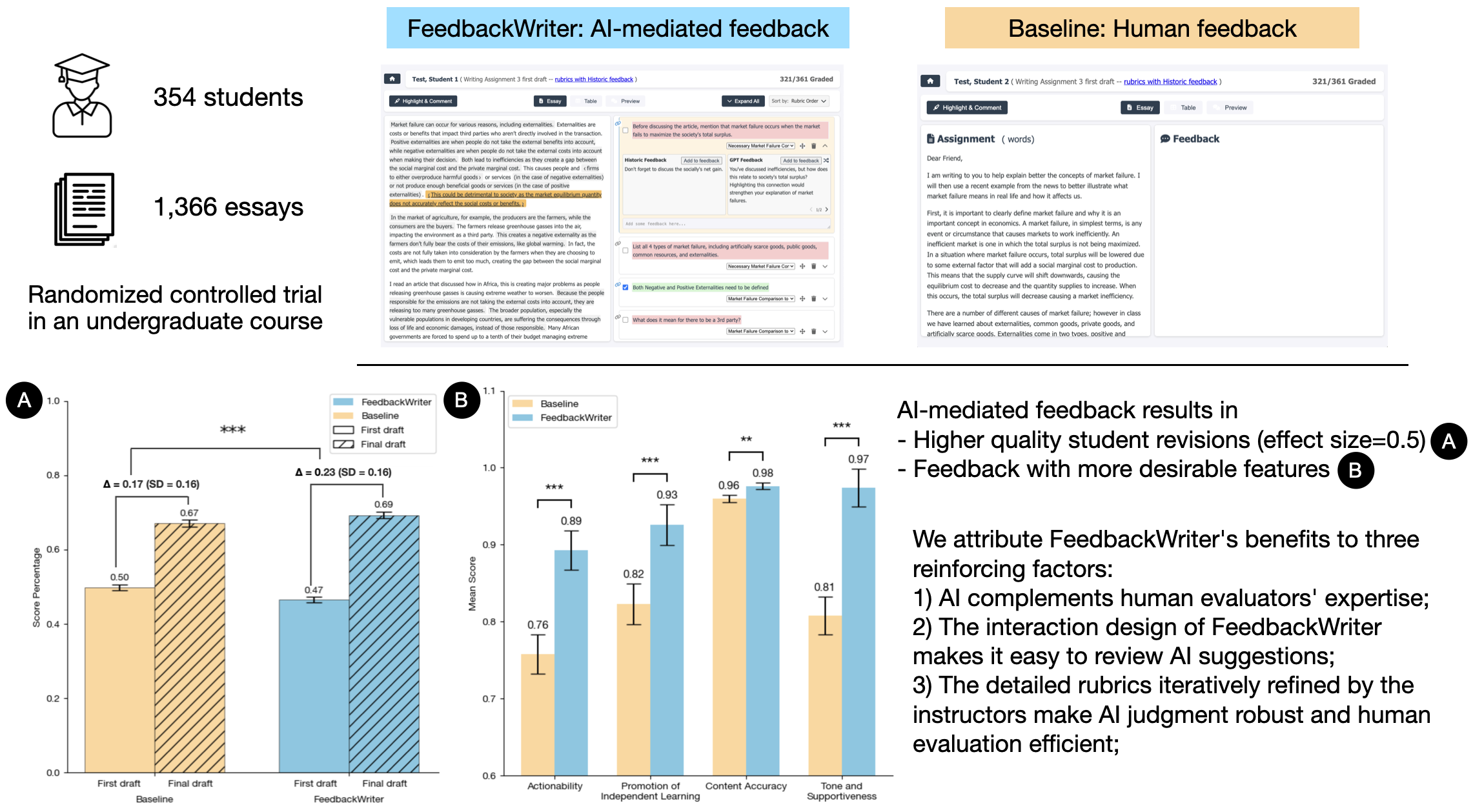}
  \caption{Our findings suggest that when making it easy for TAs to review, adopt and edit AI feedback suggestions, AI-mediated feedback authored through \textit{FeedbackWriter} has higher benefits compared to human-only feedback. Students who received AI-mediated feedback had higher quality revisions (Cohen’s d = 0.50), roughly equivalent to moving a student from the 50th to the 70th percentile. AI-mediated feedback demonstrated more desirable properties of effective feedback, such as actionability and promotion of independent learning. 
  }
  \label{fig-teaser}
\end{teaserfigure}

\maketitle

\section{Introduction}

Feedback is one of the most powerful methods to enhance learning \cite{hattie2007power, wisniewski2020power}. However, providing targeted feedback to students' open-ended written work requires substantial efforts from experts \cite{henderson2019challenges, paris2022instructors, cambre2018juxtapeer, krause2017critique}. 
Recent research has shown mixed results on using large language models (LLMs) to generate feedback in educational contexts. 
Some studies have shown promise\highlight{---}AI feedback can exhibit more desirable properties as judged by teachers or automated evaluators \cite{cao2025first, coyne2025annotating}. Only a handful of studies show improvements on student outcomes. 
For example, \cite{meyer2024using} found in a randomized trial that receiving AI-generated feedback increased middle school students' revision performance and task motivation compared to receiving no feedback. \cite{kinder2025effects} found that feedback generated by ChatGPT was more effective than pre-written static feedback in improving student's writing performance. \cite{denny2023can} showed that students rated AI-generated content to be of similar quality to peer-generated content. 
On the other hand, research has revealed frequent LLM mistakes. For example, ~50\% of math-tutoring dialogues contain mistakes \cite{gupta2025beyond}; AI-generated next-step actions reach only 52\%-70\% accuracy in TutorGym \cite{weitekamp2025tutorgym}; and 30\% of ChatGPT algebra hints were rejected for low quality \cite{pardos2023learning}. These errors in AI-generated feedback are found to negatively impact student learning outcomes \cite{bastani2025generative}.

To address AI limitations in producing pedagogical materials, one line of promising research is to explore human-AI collaboration. For example, Tutor Copilot \cite{wang2024tutor} shows exciting results that when human tutors receive AI suggestions during tutoring sessions, it leads to higher learning gains. Similarly, two studies this year \cite{gurung2025human, thomas2024improving} have found that human-AI tutoring outperforms AI tutoring alone in middle school math. However, there has been limited exploration on human-AI complementary feedback provision. 

We highlight two gaps in prior research that this project aims to address. (1) First, we explore \textit{AI-mediated feedback}, where human evaluators receive AI suggestions during feedback provision. In particular, we situate our work in \textit{knowledge-intensive essays} where students need to demonstrate accurate conceptual understanding and analyses of phenomena through writing. This may pose additional challenges for AI to make accurate judgments, different from math and argumentative writing, which are most heavily studied in the literature. (2) Second, most studies evaluate feedback quality via expert ratings \cite{chase2025genai, steiss2024comparing, barros2025large, coyne2025annotating, dai2023can} or automated analyses \cite{cao2025first, dai2023can, wang2024tutor, zhang2025friction} (e.g., annotating quality using a machine learning or LLM-based model), rather than measuring effects on learners. 
The few studies that examined student outcomes compared AI feedback versus no feedback \cite{meyer2024using} or static canned comments \cite{kinder2025effects}, which were weaker baselines. In this project, we perform the first randomized trial in a large undergraduate course (N=354) comparing the effects of AI-mediated feedback and human-only feedback on student revision quality and learning. 
Addressing these two research gaps is important to understand AI's capabilities and weaknesses in generating feedback for knowledge-intensive essays and for designing reviewer workflows that let instructors quickly triage, verify and adapt AI suggestions. 
Studying AI-mediated feedback also offers a pragmatic adoption pathway: 
students receive instructor-vetted comments, mitigating concerns about unfiltered AI outputs.

To enable AI-mediated feedback, we introduce \textit{FeedbackWriter}, a system that provides AI suggestions aligned with human evaluators' feedback provision workflow, while granting people full agency to adopt, edit, or dismiss the suggestions. 
An important input to FeedbackWriter is the essay evaluation rubrics (Table \ref{table:prompt}), which needs to be supplied by the instructors. 
For each rubric item, the system generates a feedback message through a pipeline that mirrors a human evaluator's process: (1) scanning the essay to identify relevant sentences; (2) judging whether the rubric is satisfied or missing; (3) drafting a feedback message. FeedbackWriter then anchors one AI-generated feedback message per rubric item directly in the essay, as shown in Figure \ref{fig:pipeline}. 
To support transparency and human control, the system displays intermediate outputs that users can adjust:
\begin{itemize}[nosep]
    \item Sentence highlighting: Relevant sentences are marked, with repainting options to refine anchoring. 
    \item Rubric judgment: A checkbox allows flipping the system's judgment, with color cues for satisfied vs. missing.
    \item Feedback options: Two choices are provided\highlight{---}AI-generated feedback (personalized) and instructor-supplied historic feedback (deterministic). 
\end{itemize}

Crucially, all student-facing content remains under evaluator control: final feedback boxes and scores are deliberately left empty, requiring instructors to review and confirm every message before sending to students.

\begin{table*}[htbp]
\caption{This table shows abbreviated versions of the assignment prompts, a sample of rubrics for each assignment, student mistakes, and AI-generated feedback. The writing assignments are knowledge-intensive essays, in which students are required to demonstrate an accurate understanding and application of economic concepts. The rubrics are specifically designed to capture knowledge deficiencies instead of writing prose.}
\begin{tabular}{|p{0.26\linewidth}|p{0.20\linewidth}|p{0.22\linewidth}|p{0.25\linewidth}|}
\hline
Abbreviated Version of the Assignment Prompt &
Example Rubric &
Student Mistake &
AI Feedback \\ \hline

\multirow{2}{=}{Given the rise in preference to oat milk, analyze one of the two policies: either a price floor for almond milk to support the almond milk market, or a tax on agricultural water use to reduce water usage. Explain its expected effects, trade-offs, and impact on society.}
&
Identify that almond milk and oat milk are substitutes in consumption.
&
Student only analyzed the influence of preference change on the oat milk market; did not discuss the almond milk market.
&
What is the relationship between almond milk and alternative milk in terms of consumer choice? How might this relationship affect market dynamics?
\\ \cline{2-4}

&
Explain that both the consumer and the producer will bear the burden of the tax.
&
Student only analyzed the influence of the tax on consumers, but did not consider suppliers.
&
Who is bearing the burden of the tax? Think about how the tax affects both sides of the market.
\\ \hline

\multirow{2}{=}{Identify a recent news article that contains a market failure, explain the market failure within it, and propose a solution.}
&
Propose and expand well on a thoughtful solution to the negative externality in the article.
&
The student did not expand on the policy requirement, such as the efficient size.
&
How can the efficient size of a tax be determined, and why is this important for addressing the negative externality?
\\ \cline{2-4}

&
Before discussing the article, mention that market failure occurs when the market fails to maximize society's total surplus.
&
Student did not differentiate between the private inefficiency and society's total surplus.
&
You've discussed inefficiencies for consumers, but how does this relate to society's total surplus?
\\ \hline
\end{tabular}
\label{table:prompt}
\end{table*}

We deployed FeedbackWriter in an Introduction to Economics class (ECON101) at a public R1 university in the United States, with 354 students and 11 teaching assistants (TAs), who served as evaluators of student writing. The TAs were selected through a competitive process and participated in training activities such as grade norming and group discussions of evaluation criteria. This course had 2 writing assignments per semester with detailed rubrics that requires the students to apply and demonstrate their economic knowledge, as shown in Table \ref{table:prompt}. For each assignment, students submitted a first draft, received feedback, and then submitted a final revised draft. 
At the start of the semester, students were randomly assigned to one of two conditions. 
For the first assignment, half received traditional human feedback, while the other half received AI-mediated feedback in which TAs received LLM suggestions when providing feedback. In the human feedback condition, TAs have access to the same rubrics and historic feedback as in the FeedbackWriter condition. 
The conditions were reversed for the second assignment. Over the semester, TAs used FeedbackWriter to provide feedback on 1,366 essays. 
Here is a summary of the findings:
\begin{enumerate}[nosep]
    \item Students who received AI-mediated feedback produced significantly higher-quality revised drafts than students who received human-only feedback (Cohen’s d = 0.50), roughly equivalent to moving a student from the 50th to the 70th percentile. The effect is larger when the TAs adopted more AI-suggested constructive feedback. 
    \item Having a higher-quality revised draft is associated with higher learning (measured by a post-test following students' submission of the revision). However, there is no statistical difference between the two conditions on the post-test scores. 
    \item An automated analysis of feedback quality revealed that AI-mediated feedback outperformed human-only feedback, offering significantly greater actionability and stronger support for independent learning. 
    \item TAs found the AI suggestions to be generally accurate and helped them uncover overlooked errors in student essays, understand rubrics, and frame feedback. They also appreciated that FeedbackWriter made it straightforward to detect and correct AI-generated mistakes.
\end{enumerate}

This work makes the following contributions.
We introduce FeedbackWriter, a system that supports the provision of AI-mediated feedback  \highlight{through} visualizing multi-turn feedback anchored in long knowledge-intensive essays and supporting the human evaluators to easily adopt, edit or dismiss AI suggestions. 
We then present the first randomized trial comparing the effects of AI-mediated feedback and human-only feedback on student revision quality and learning. We present promising results that students who receive AI-mediated feedback produce higher quality revisions.

\section{Related Work}

\subsection{Writing-to-Learn Assignments}
Writing-to-Learn (WTL) is a widely adopted instructional approach where students learn disciplinary ideas through cycles of writing, feedback, and revision \cite{finkenstaedt2023portrait, reynolds2012writing, keys1999using, gunel2007writing}. Originating in the humanities, WTL has been successfully translated to STEM classrooms, including nursing \cite{schmidt2004evaluating}, psychology \cite{nevid2012writing, gingerich2014active}, and mathematics \cite{countryman1992writing, bicer2018impact, tong2009integrating}. Through applying concepts and externalizing reasoning, WTL consolidates conceptual knowledge \cite{keys1999using, reynolds2012writing} and promotes disciplinary reasoning and problem-solving skills \cite{quitadamo2007learning, tong2009integrating}. WTL assignments are typically knowledge-intensive, emphasizing the accuracy of ideas and quality of analysis rather than \highlight{prose} \cite{prain1996writing, moon2018writing}. 
\highlight{Effective feedback is central to these cycles: it provides targeted instruction to help students recognize misconceptions \cite{kluger1996effects, pan2023prequestioning}, 
scaffold their reasoning process \cite{sadler1989formative, hattie2007power, nicol2006formative}, 
improve learning efficiency \cite{asher2024practice, carpenter2017effect}, and guides revision toward intended goals \cite{sadler1989formative, hattie2007power, kluger1996effects}}. 
This work investigates AI-mediated feedback on knowledge-intensive essays and compares its effects to human-only feedback on students’ revision and learning.
\subsection{Existing Work on Using AI to Generate Feedback}
Despite AI's promise of enabling personalized comments at scale \cite{almegren2024evaluating,latif2024fine, tran2025enhancing}, research on AI-generated feedback showed mixed results.

First, several studies report benefits---e.g., AI feedback outperforming feedback written by novices \cite{jacobsen2025promises, usher2025generative}, students often perceive AI feedback as useful \cite{escalante2023ai, jia2024assessing}, and receiving additional AI feedback leads to improved learning outcomes \cite{thomas2025llm}. 

However, much of the positive evidence come from relatively weak comparisons (e.g., no feedback or novice feedback). When compared to expert feedback directly, AI feedback is less preferable due to misalignment with teaching goals \cite{jia2024llm} and containing incorrect or irrelevant comments \cite{jayawardena2025dental}. For instance, when blind to the sources, students rejected more than 40\% of the AI feedback on their academic writing, while the rejection rate for instructors' feedback was less than 20\% \cite{lu2024can}.

Second, most prior work focused on short-answer questions with few knowledge requirements. With comprehensive correct answers and relevant domain knowledge provided, LLMs produce concrete and factually correct feedback \cite{wan2024exploring, matelsky2023large, jacobsen2025promises, usher2025generative}. In contrast, for longer essays, studies documented recurring hallucinations and mistakes \cite{lu2024can, gupta2025beyond}. LLMs frequently struggle to identify errors within a student's specific context \cite{Seler2025TowardsAF} and tend to prioritize grammar and syntax mistakes, whereas expert feedback often addresses ``higher-order'' and knowledge-level concerns \cite{tran2025enhancing}, especially when rubrics are not explicitly articulated \cite{Almasre2024DevelopmentAE}. Beyond essay length, knowledge-intensive assignments introduce additional challenges. Empirical studies reported frequent AI mistakes in domain-knowledge, especially in STEM domains \cite{gupta2025beyond, pardos2023learning, weitekamp2025tutorgym, liu2025llms}, with a 30\% to 50\% failure rates in math \cite{pardos2023learning, gupta2025beyond}, programming \cite{silva2025assessing} and dentistry \cite{jayawardena2025dental}. 

Third, recent work also shows that achieving high-quality AI feedback often rely on careful, iterative prompt design \cite{thomas2025llm, lu2025exploring, lu2024can}. Effective setups typically combine general prompting techniques (e.g., few-shot prompting) \cite{wan2024exploring, jayawardena2025dental, thomas2025llm} with explicit feedback guidelines \cite{lu2025exploring} and rich task context (e.g., role, audience, goals) \cite{wan2024exploring}. To approach expert-like performance, systems frequently require expert-provided artifacts such as example feedback \cite{wan2024exploring}, detailed rubrics \cite{lu2025exploring}, and domain knowledge \cite{jacobsen2025promises, jayawardena2025dental}. Without these supports, AI feedback is more likely to be generic \cite{jia2024llm}, reveal solutions \cite{macina2023mathdial}, or drift from instructional goals \cite{jacobsen2025promises, usher2025generative}.

\subsection{Human-AI Collaborative Feedback Provision}
Education research has increasingly emphasized human-AI collaboration over full automation, leveraging the complementary strengths of AI and human. AI supports experts by providing guideline-aligned suggestions \cite{wang2024tutor, wan2024exploring}, offering language-level support \cite{lu2023readingquizmaker, pahi2024enhancing}, and streamlining repetitive, labor-intensive tasks, such as monitoring student progress \cite{zhang2023vizprog, yang2025spark}. Human experts remains ``in the loop'' for guidance and oversight.


Building on the mixed results on AI-generated feedback, recent work examines human–AI collaborative feedback provision \cite{lu2024can, yang2025exploring, usher2025generative}. Studies in math and coding reported promising findings that the AI-mediated feedback outperforms AI-generated feedback in technical quality and identifying next steps \cite{pahi2024enhancing}, resulting in an improved learning efficiency \cite{gurung2025human}. AI-mediated feedback also outperformed human-only feedback on identifying language errors \cite{yang2025exploring}, gaps in students’ work, and providing encouraging guidance \cite{pahi2024enhancing}, leading to greater learning gains \cite{wang2024tutor, thomas2024improving}.

However, it remains unclear how to enable effective human-AI collaborative feedback on knowledge-intensive essays. Additionally, it remains underexplored how students respond to AI-mediated feedback in comparison to expert feedback.
We address these gaps by (1) engaging in knowlege engineering efforts to surface knowledge requirements and represent them as rubric items and (2) designing an interface that provides AI suggestions aligned with experts’  feedback provision workflows. Then we compare student uptake on the AI-mediated feedback versus human-only feedback through a randomized controlled trial.

\subsection{Metrics to Evaluate Feedback Quality}

A large body of work identifies key dimensions of high-quality feedback, including length, actionability, justification, specificity, and supportiveness, \cite{ngoon2018interactive, krause2017critique,Steiss_Tate_Graham_Cruz_Hebert_Wang_Moon_Tseng_Warschauer_Olson_2024, Kakarla_Thomas_Lin_Gupta_Koedinger_2024}, which correlate with perceived helpfulness and uptake \cite{krause2017critique, zhang2025friction}. 
Tutoring frameworks similarly define effective feedback strategies, especially in STEM contexts \cite{Lepper_Woolverton_2002, chhabra2022evaluation, kakarla2024using, wang2024tutor}. For example, the INSPIRE model characterizes effective tutoring as intelligent, nurturant, Socratic, progressive, indirect, reflective, and encouraging \cite{Lepper_Woolverton_2002}, and recent work applied this model to evaluate tutors on criteria such as being process-focused, motivating, indirect, immediate, and accurate \cite{kakarla2024using}. Within this line of work, practices like indirect hints and Socratic questioning are often treated as observable markers of higher-quality feedback \cite{wang2024tutor}.
In writing, feedback was categorized by type (summary, praise, problem, solution) and focus (low prose, high prose, substantive content) \cite{Patchan_Schunn_Correnti_2016}, with high-prose or substance-focused feedback more likely to improve revision quality. In this work, we categorize feedback by type following \cite{Patchan_Schunn_Correnti_2016}, and evaluate feedback quality using a set of dimensions synthesized from prior feedback-quality literature and the INSPIRE tutoring model.

When studying the quality of AI-generated feedback, most evaluations of feedback quality stop at expert ratings or automated analysis, without measuring downstream student uptake or learning \cite{dai2023can, Steiss_Tate_Graham_Cruz_Hebert_Wang_Moon_Tseng_Warschauer_Olson_2024}.
\highlight{The limited research on learning typically compared AI-generated feedback to no-feedback \cite{meyer2024using}, static, pre-written feedback from experts \cite{escalante2023ai, alnemrat2025ai, kinder2025effects, pardos2024chatgpt, pardos2023learning}, or student-sourcing feedback \cite{denny2023can}.}
Thus, it remains unclear how AI-mediated feedback --- where experts collaborate with AI --- compares to experts-provided feedback in uptake, learning, and revision.
To address this gap, we deployed FeedbackWriter in a real classroom, and compared AI-mediated and human-provided feedback on both the quality of feedback and its impact on students.

\section{Formative Study}
We conducted a formative study to understand (1) TAs’ natural workflow for essay feedback, (2) challenges in their current practice, and (3) perspectives toward using AI-generated suggestions. 
Then, we summarize the design requirements for developing technologies to support the process.

\subsection{Participants and Procedure}
We recruited five experienced TAs with 1-5 semesters of experience from an Introduction to Economics class (ECON101) in Fall 2024 and followed them throughout the semester in 20 one-hour, think-aloud sessions on Zoom. We describe the course context in detail in Section \ref{section:course_context}. Participants received a \$25 gift card per session.

To ensure authentic comparison, TAs first provided feedback on the essays as usual before the think-aloud sessions. During the sessions, participants were shown samples of AI-generated comments, which were generated in advance by a pipeline \xinyi{powered by GPT-4o}, which took the grading rubrics as input and generated a comment for each rubric. The generated comments were embedded as comments in a Word document alongside the student essay. Participants reviewed the essay and the comments through the Word document, and compared these AI-generated comments with their handwritten ones during the think aloud session. 
We also explicitly asked them to share their challenges. 

\highlight{The interview recordings were transcribed and analyzed using affinity diagrams \cite{lucero2015using}. Two authors open-coded 5 sessions together and created the initial set of interpretation notes to create an affinity diagram. The two researchers then coded the rest of the transcripts independently. They then met to discuss and resolve the conflicts, incorporate new interpretations, and update the affinity diagram.}

\subsection{Results} 
\subsubsection{The feedback provision process is rubric-driven and time-consuming.}
TAs relied heavily on rubrics when giving feedback. After reading an essay to grasp its overall flow, they evaluated each rubric separately: skimming the essay to locate relevant sentences, assessing its adequacy for the rubric, and drafting feedback to guide revisions. This process is time-consuming, especially the searching step. \highlight{P2 noted, \textit{``It can get fatiguing ..., especially with so many papers ... and make sure I'm giving everyone explicit comments.''}} \highlight{This aligns with prior work that rubrics are key mechanisms for consistency and fairness in learning assessments. \cite{huba2000learner, stevens2023introduction, yuan2016almost}}

\subsubsection{Historic feedback supports comment phrasing but reduces personalization.}
Despite training, TAs expressed uncertainty about phrasing constructive feedback for diverse submissions, especially asking effective guiding questions without revealing answers. \highlight{For example, P4 mentioned, \textit{``I want to do (ask probing questions) but sometimes it's just hard.''}} Participants often drew heavily on historic feedback or instructor examples, reusing comments across students. They emphasized the cognitive and time pressure of producing personalized comments at scale, given the volume of essays and short turnaround time. As P4 explained, \textit{``The main time is writing those comments ... individually for everyone.''}

\subsubsection{Coordinating multiple reference sources disrupts workflow.}
TAs draw on multiple resources when giving feedback. Beyond the essay and rubrics, they consult meeting notes (P1), historic feedback (P2, P4, P5), and sometimes their comments on other students’ essays (P1, P3) to ensure consistency. This process requires navigating 3-4 browser tabs, which TAs found frustrating and prone to disrupting their progress (P2).

\subsubsection{TAs' feedback lacks coherence due to challenges in noticing all relevant sentences and discrepancies in their understanding of the rubrics.}
During think-aloud sessions, several participants realized they had overlooked student mistakes \highlight{(P2, P5)} or evidence of satisfying a rubric \highlight{(P1, P4)}. They expressed concern about the fairness, and noted inconsistent interpretations of rubric expectations, especially around the required depth of analysis, despite shared rubrics and regular staff discussions.
For a criterion such as ``a thoughtful and well-reasoned solution,'' P5 expected a thorough explanation of why the solution would work, whereas P2 only expected simply naming a solution. Discrepancies often arose at rubric boundaries, where imprecise wording (P4) or responses that were close but incomplete (P5) made judgments difficult.

\subsubsection{AI suggestions could improve feedback comprehensiveness and quality, but may have hallucinations and additional reading loads from AI suggestions.}
Participants envision AI-generated suggestions as helpful for surfacing overlooked points, and facilitating their feedback writing with more praise (P1, P5), guiding questions (P2), and explanations of the mistakes (P4, P5). TAs lack the bandwidth and time to polish their own feedback to this standard.
However, participants shared worries about hallucinations and missing key points. They require agency over AI suggestions, emphasizing a verification-first approach (P1-P3, P5): making judgment \highlight{before reading} AI suggestions as supplements. 
Moreover, as feedback provision is highly cognitively demanding, participants shared concerns that comprehending and evaluating AI suggestions could add cognitive load rather than reduce it (P2, P3).

\subsection{Design Goals}
Based on the formative study findings, we summarized the following design goals (DGs) for creating systems to support the instructors in creating scaffolding questions:
\begin{itemize}

    \item[\textbf{DG1}] \textbf{Facilitate feedback provision while adding minimal cognitive load.} The system should stick to the rubrics and help users 1) comprehensively identify relevant sentences; 2) provide consistent judgment on rubric fulfillment; and 3) construct effective feedback aligning with guidelines. 
    \item[\textbf{DG2}] \textbf{Enable easy navigation between relevant information sources.} The system should lower the effort of moving among the essay, rubrics, historic feedback, and the students' first draft with its feedback, when grading the final draft.
    \item[\textbf{DG3}] \textbf{Make AI suggestions easy to understand and inspect, and AI mistakes easy to recognize and correct, while keeping users in control.} Each suggestion should be tied to the specific rubric, and provide a clear rationale that is esay to inspect without adding reading burden, making mistakes easy to recognize and correct. Users should have full control over when to adopt, edit, or dismiss AI suggestions. 
\end{itemize}

\section{FeedbackWriter}

\begin{figure*}
    \centering
  \includegraphics[width=\textwidth]{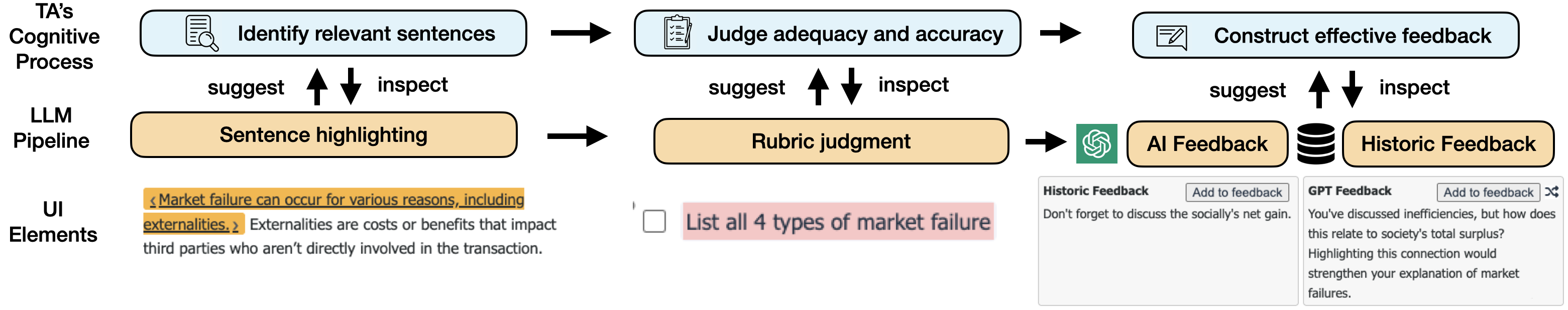}
  \caption{The design of FeedbackWriter follows TAs' natural workflow of feedback provision. AI suggestions are provided for each rubric and are surfaced through corresponding UI elements, including 1) identifying relevant sentences (sentence highlighting); 2) judging whether the rubric is adequately and accurately addressed (rubric judgments); and 3) constructing effective feedback (AI feedback and historic feedback).}
  \Description{Three-row diagram showing the alignment of TAs’ cognitive process with an LLM pipeline and example UI. The top row shows three steps in "TA’s Cognitive Process" left to right: Identify relevant sentences → Judge adequacy and accuracy → Construct effective feedback. The middle row shows the "LLM Pipeline" mirrors these steps with modules: Sentence highlighting → Rubric judgment → AI Feedback / Historic Feedback. Between the top and middle rows, two-way arrows labeled "LLM suggest" and "TA inspect" indicate that TAs review the system’s suggestions at each stage. The bottom row shows the "UI Elements" examples for each step: highlighted text in a student response, a rubric checklist item (“List all 4 types of market failure”) in red background color with an unchecked checkbox, and side-by-side feedback panels labeled Historic Feedback and GPT Feedback with an “Add to feedback” option.}
  \label{fig:pipeline}
\end{figure*}

\begin{figure*}[htbp]
    \centering
    \includegraphics[width=\linewidth]{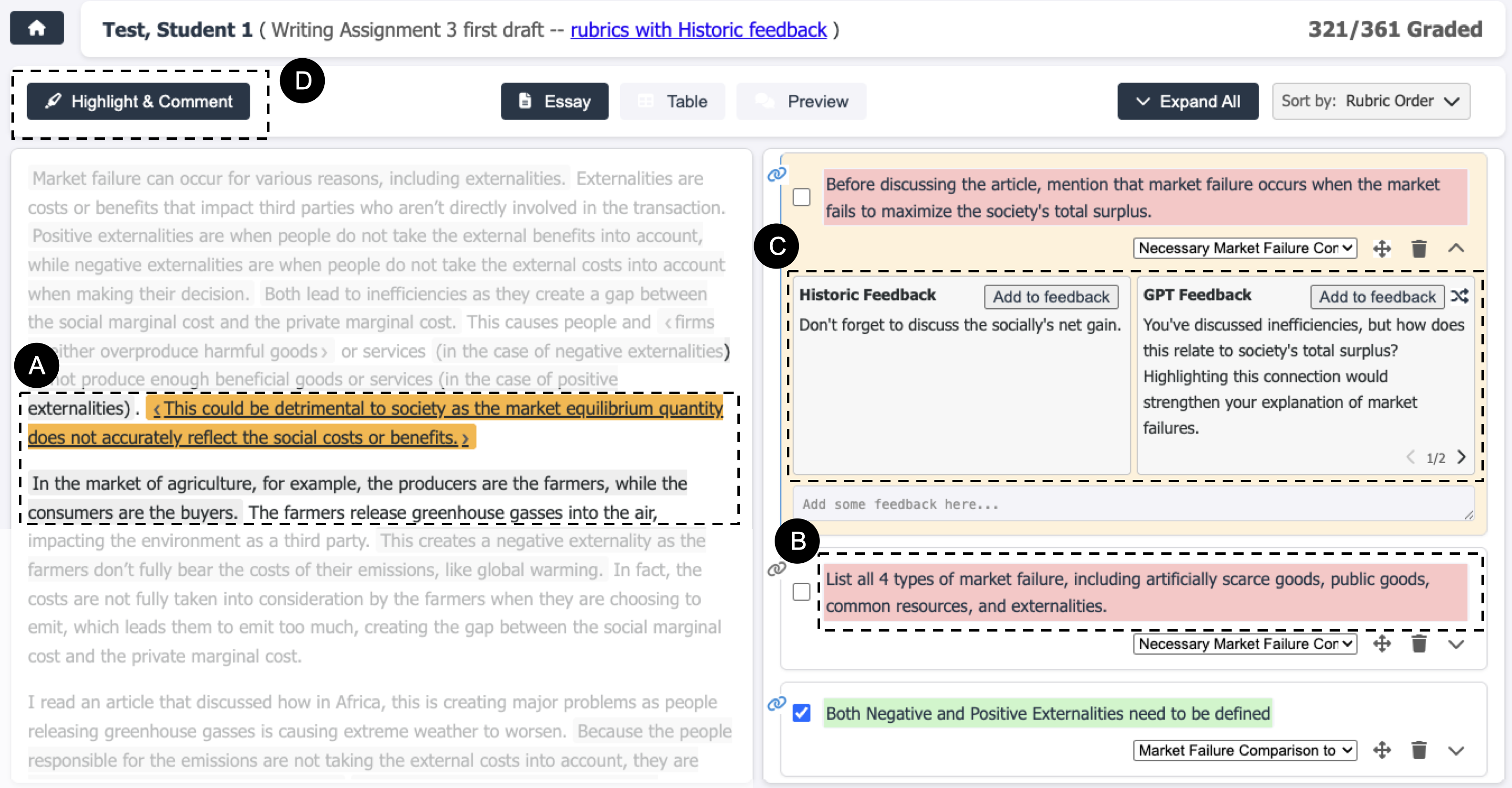}
    \caption{FeedbackWriter interface contains an essay panel (left) and a feedback panel (right). The essay panel displays the student essay, where the relevant sentences associated with each feedback are highlighted (A). The feedback panel displays AI-generated comments organized by rubrics. Each rubric item has its own feedback box, including the rubric, rubric judgment with color cues (B) and two feedback suggestions (C). Users can also manually add additional in-text comments (D).}
    \Description{Annotated screenshot of a web-based interface. The screen is split into two panels. Left panel: the student’s essay text; a sentence is highlighted in orange and boxed. The box is labeled as A. Top-left toolbar: a "Highlight & Comment" button is boxed and labeled D. Right panel: multiple cards each containing a feedback item for a rubric. Each card contains a rubric item on top with a checkbox, a tool box, two feedback suggestions and a textarea with placeholder text "Add some feedback here...". Two rubrics are highlighted in red background color and labeled B. One rubric is highlighted in green with the checkbox checked. Feedback suggestions in one feedback box is labeled C, with side-by-side cards for Historic Feedback and GPT Feedback.}
    \label{fig:essay_view}
\end{figure*}

Based on the findings and the design goals from the formative study, we developed FeedbackWriter, an integrated workspace for feedback provision that presents AI suggestions aligned with human evaluators' feedback provision workflow, while granting users full agency to adopt, edit, or dismiss the suggestions. 
FeedbackWriter has two main components: 1) a pipeline for feedback generation and 2) an interactive interface to facilitate human-AI collaboration. 
\subsection{Cognitive-Aligned Feedback Pipeline and Interactive Designs}
\label{pipeline}

To achieve (DG1), an important design consideration of FeedbackWriter is to align AI's feedback generation process with that of the human evaluators. 
For each rubric item, the system 
(1) identifies relevant sentences; (2) judges whether the rubric is satisfied; (3) drafts feedback (Figure \ref{fig:pipeline}). The FeedbackWriter interface presents the student essay alongside a dedicated feedback panel (Figure \ref{fig:essay_view}). To support transparency and human control (DG3), the system displays intermediate AI outputs from all steps for user inspection, and provides interactions for adjustment. All subtasks are implemented as GPT-4o calls with temperature 0.05.

\textbf{Step 1: Identify relevant sentences in the essay that address the rubric.}
For each rubric item, the model is prompted to provide a comprehensive list of sentences where the student attempts to address the rubric, to support a judgment, following the extractive QA practice suggested in \cite{rajpurkar2018know}.

These highlights are visualized on the essay, which users can modify 
by repositioning them (drag-to-select) or fine-tuning their boundaries with draggable cursors.

\textbf{Step 2: Make a judgment on whether the rubric is satisfied.}  
The model applies explicit rules: label the rubric as met only when all required elements, and terms are present and accurate \highlight{without mistakes. The model is required to provide a rationale before the judgment \cite{kojima2022large}.} 
\highlight{The interface exposes this judgment through a checkbox with color cues. Users can flip the decision if the model misapplies the rubric.} Changing the label also automatically switches the AI-generated message between praise (when a rubric is met) and a constructive feedback message (when a rubric is missing). 

\textbf{Step 3: Generate a feedback message to help students satisfy the rubric in the revision.} 
The feedback generation follows strategies for effective feedback, including 1) using specific and localized language to pinpoint mistakes or achievements; 2) praising when the rubric is satisfied; and 3) posing Socratic questions that highlight gaps without revealing the answer. Specifically, to preserve students' individual learning opportunities, the feedback is required to be framed as ``hints'' that direct students toward the right questions or resources. The prompt also includes three instructor-provided exemplar questions for few-shot guidance \cite{brown2020language}.

\begin{figure}
    \centering
    \includegraphics[width=\linewidth]{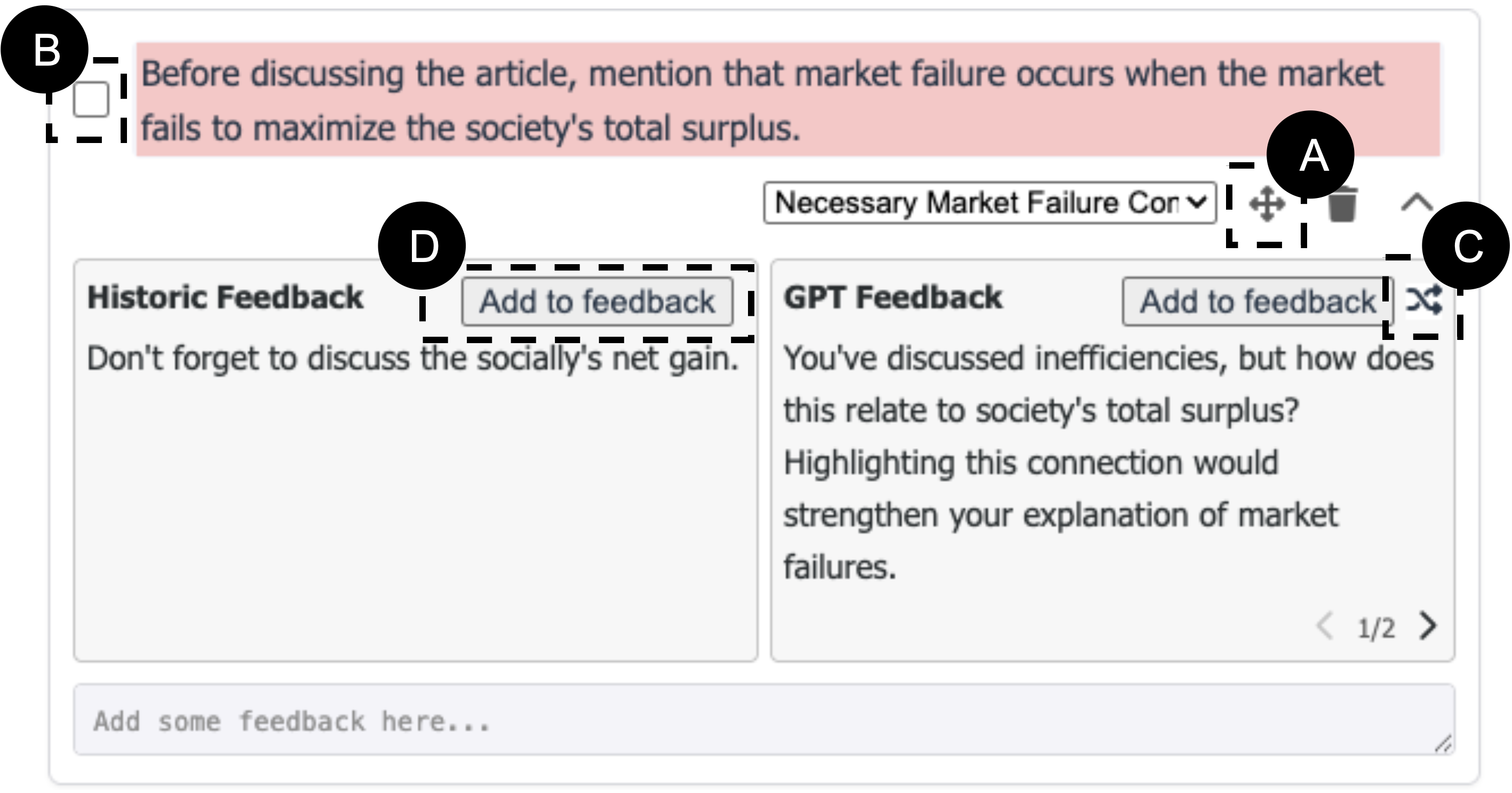}
    \caption{In FeedbackWriter, users can use and adjust all AI suggestions conveniently. Users can use (A) the "reposition" button to modify the highlight, click (B) to flip the judgment, and click (C) the "regenerate" button to get another suggestion. The feedback boxes are empty by default, and users can click (D) to adopt either feedback suggestion.}
    \Description{Annotated close-up of a feedback box on the FeedbackWriter interface. At the top, the rubric is highlighted in a red background color, with a checkbox at the left of the rubric is boxed and labeled B. On the right side of the comment row, the tool box contains a rubric-category dropdown, an anchor icon, and a delete icon. The anchor icon is boxed and labeled A. Below, two side-by-side panels compare Historic Feedback (left) and GPT Feedback (right). The Historic Feedback panel says "Don’t forget to discuss the socially’s net gain." and the GPT Feedback panel provides a longer suggestion about connecting inefficiencies to society’s total surplus. Both cards has an "Add to feedback" button boxed on the right-top corner, boxed and labeled D. In the GPT Feedback area, an exchange icon is boxed and labeled C. A text field with placeholder text "Add some feedback here ..." appears at the bottom.}
    \label{fig:AI_modification}
\end{figure}

The interface provides two feedback options side-by-side: personalized AI-generated feedback and instructor-provided historic feedback (deterministic). This layout design is inspired by the formative study, where TAs often consult prior notes with historic feedback when crafting new feedback messages. To discourage overreliance on AI, the feedback boxes are empty by default (Figure \ref{fig:AI_modification}). Instructors can insert a suggestion into the final feedback box\highlight{, edit it, regenerate an AI suggestion based on the ``relevant sentences'' and the ``judgment'', or write their own message.}

In our experiments with the feedback generation pipeline, we found that clarity of the rubrics was essential. During implementation, we had iterative consultations with the course instructors to improve the rubrics. Table \ref{table:modifications} shows example changes to clarify the rubrics for LLMs. Specifically, as TAs have the course-specific context and domain knowledge that LLMs lack, additional clarifications on the context and terminology are included to mitigate hallucinations. Since our pipeline is designed to adhere closely to rubric items, we also specified the depth of explanation required and provided acceptable alternatives to introduce reasonable leniency for valid but diverse student answers. \highlight{For example, the rubric item ``Perform a thorough welfare analysis'' requires discussion on all categories of welfare taught in class, including consumer surplus, producer surplus, and deadweight loss. TAs would know the details from the course content, but without such contextual knowledge, LLMs might consider a discussion on deadweight loss to be sufficient. The improved rubric also made explicit the expectation to analyze both the existence and the magnitude of deadweight loss.}

\begin{table*}[htbp]
\caption{Example changes made to adapt the originally TA-facing rubrics for LLM.}
\begin{tabular}{|p{0.23\linewidth}|p{0.17\linewidth}|p{0.33\linewidth}|p{0.12\linewidth}|}
\hline
Modifications & Original Rubrics & Updated Rubrics & \highlight{Assignment} \\ \hline

Include course and assignment context (e.g., abbreviations and examples discussed in class)
&
Discuss policy choice sets for externalities as discussed in class
&
List at least 2 possible solution policies, including legal standards, Pigouvian taxes, tradable permits, standards, and production quotas
&
Writing Assignment 2
\\ \hline

Include economic term explanations
&
Point out why the public good in the article has free-riders
&
Point out why the public good in the article has free-riders, i.e., people benefit from the good without paying for it
&
Writing Assignment 2
\\ \hline

Add localization terms to specify where the rubric applies
&
Identify the decision-makers in the example
&
When discussing the article, identify the decision-makers in the market who did not produce at the ideal social amount
&
Writing Assignment 2\\ \hline

Specify the expected depth of the explanation
&
Perform a thorough welfare analysis
&
Explain why deadweight loss exists, and mention that it is quite large given that the government purchased the excess
&
Writing Assignment 1\\ \hline

Include acceptable alternatives
&
Explain that farmers need water to produce
&
Demonstrate understanding that farmers demand water, or analyze the influence on farmers as consumers of water
&
Writing Assignment 1
\\ \hline

\end{tabular}
\label{table:modifications}
\end{table*}

\subsection{UI Improvements to Enhance Usability}
\highlight{Over a four-month iterative cycle, the team met weekly with the instructor to review progress and gather feedback, and ran pilot tests with TA on representative tasks. Their feedback guided further improvement on the usability}. Notable improvements include:

\textbf{Bi-directional anchoring \& synchronized scrolling:} Because the essay and feedback panels often differ in length, the interface provides synchronized, two-way navigation. Selecting a feedback box scrolls the essay to its anchored sentences (exposing the basis for the AI judgment); selecting a highlighted span scrolls the feedback panel to the associated feedback boxes.

\begin{figure*}[htbp]
    \centering
    \includegraphics[width=\linewidth]{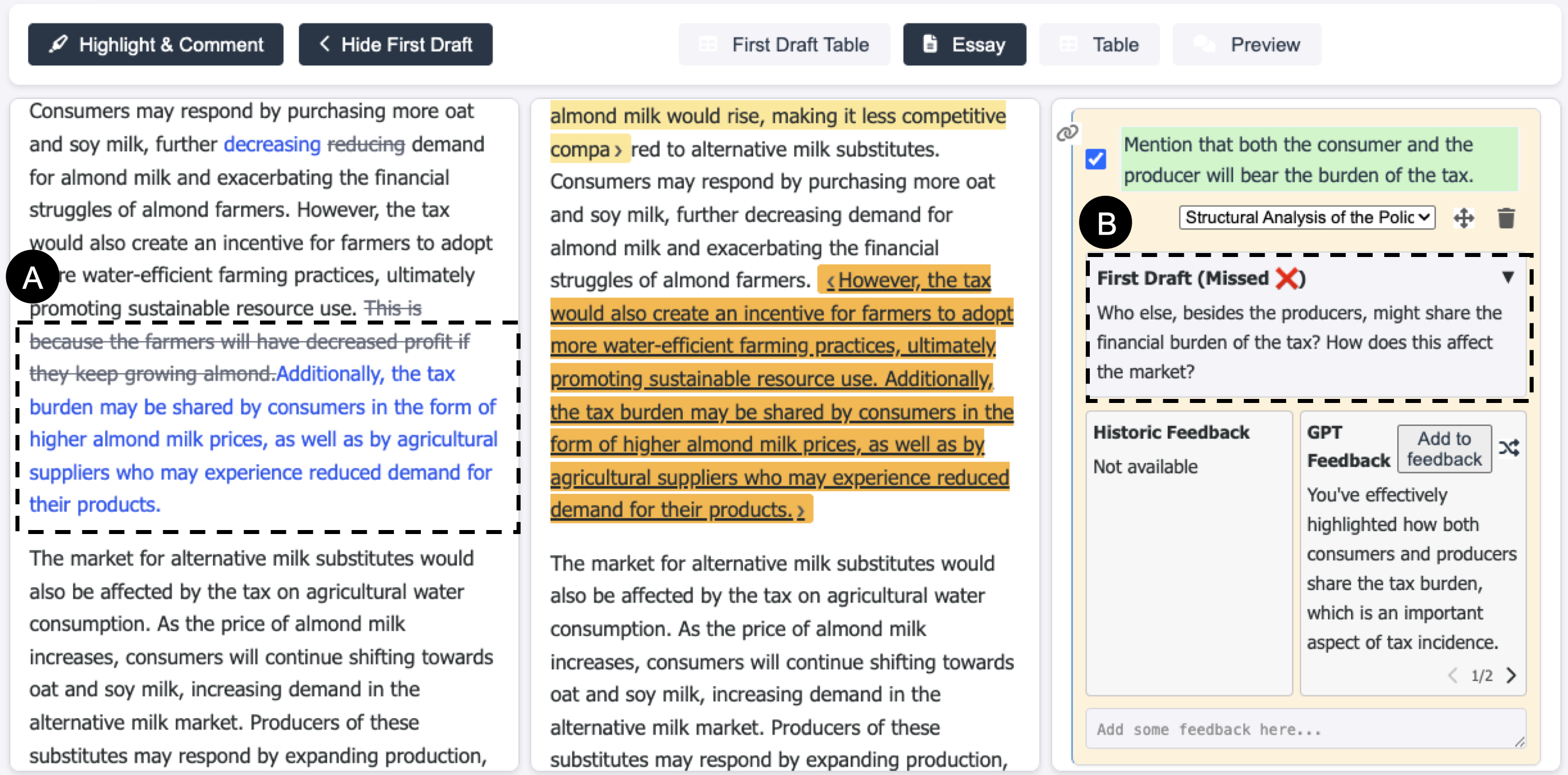}
    \caption{FeedbackWriter further supports the TAs on grading the final drafts in two ways. (A) a collapsible "diff" panel that visualizes students' additions and deletions between the first and final drafts; (B) feedback from the first draft on the same rubric item is shown in the feedback box.}
    \Description{FeedbackWriter further supports the TAs on grading the final drafts in two ways. First, a collapsible "diff" panel is added to the left of the interface, which visualizes students' additions and deletions between the first and final drafts. Second, feedback from the first draft on the same rubric item is shown in each feedback box in the feedback panel.}
    \label{fig:final_draft_view}
\end{figure*}

\textbf{Revision comparison \& prior-feedback context:} 
Since TAs revisit the first draft and their own comments when evaluating revised submissions, FeedbackWriter adds a collapsible ``diff'' panel visualizing \highlight{changes} between drafts (Figure \ref{fig:final_draft_view}); each rubric’s feedback box also surfaces the TA’s first-draft comment for quick reference.

\textbf{Additional, non-rubric feedback:} To support additional personalized comments outside the rubric, users can insert freeform feedback anchored to any essay span.

\textbf{Synchronized editing across views:} A Table view aggregates all rubric judgments and feedback, which is synchronized with the Essay view for rapid synthesizing and scoring.  
A Preview tab shows the final student-facing feedback.


\subsection{Implementation}
FeedbackWriter is implemented as a full-stack web application using React.js \cite{react} frontend and Django \cite{django} backend frameworks. We adopted the OpenAI chat API with GPT-4o model for the pipeline \cite{openai}. The web app is deployed through Google Cloud Platform \cite{gcp}.

\section{Deploying FeedbackWriter: A Randomized Trial in a Large Undergraduate Economics Class}
We performed an IRB-approved randomized controlled trial in a large undergraduate economics course at University of Michigan with 354 students and 11 teaching assistants (TAs). The study was conducted between January and April in 2025. We will refer to this class as ECON101. Students were randomized to receive either AI-mediated feedback authored with FeedbackWriter or human-only feedback produced with a baseline tool that mirrored FeedbackWriter’s features\highlight{---}anchored comment boxes, Table and Preview views, and a spreadsheet of detailed rubrics and historical feedback, but without the AI suggestions. We investigated the following research questions:

\begin{itemize}
    \item RQ1: How does AI-mediated feedback affect students' revision and learning compared with human-only feedback? 
    \item RQ2: How does feedback quality differ between the FeedbackWriter (AI-mediated) and baseline (human-only) conditions? 
    \item RQ3: How do TAs engage with FeedbackWriter's AI features while authoring feedback? 
    \item RQ4: When do TAs accept, edit, or dismiss AI suggestions, and what considerations drive these choices?
    \item RQ5: How do TAs perceive FeedbackWriter and the human-AI teaming approach in their feedback provision process?
\end{itemize}

\subsection{Course Context} \label{section:course_context}
\begin{figure}[htbp]
    \centering
    \includegraphics[width=\linewidth]{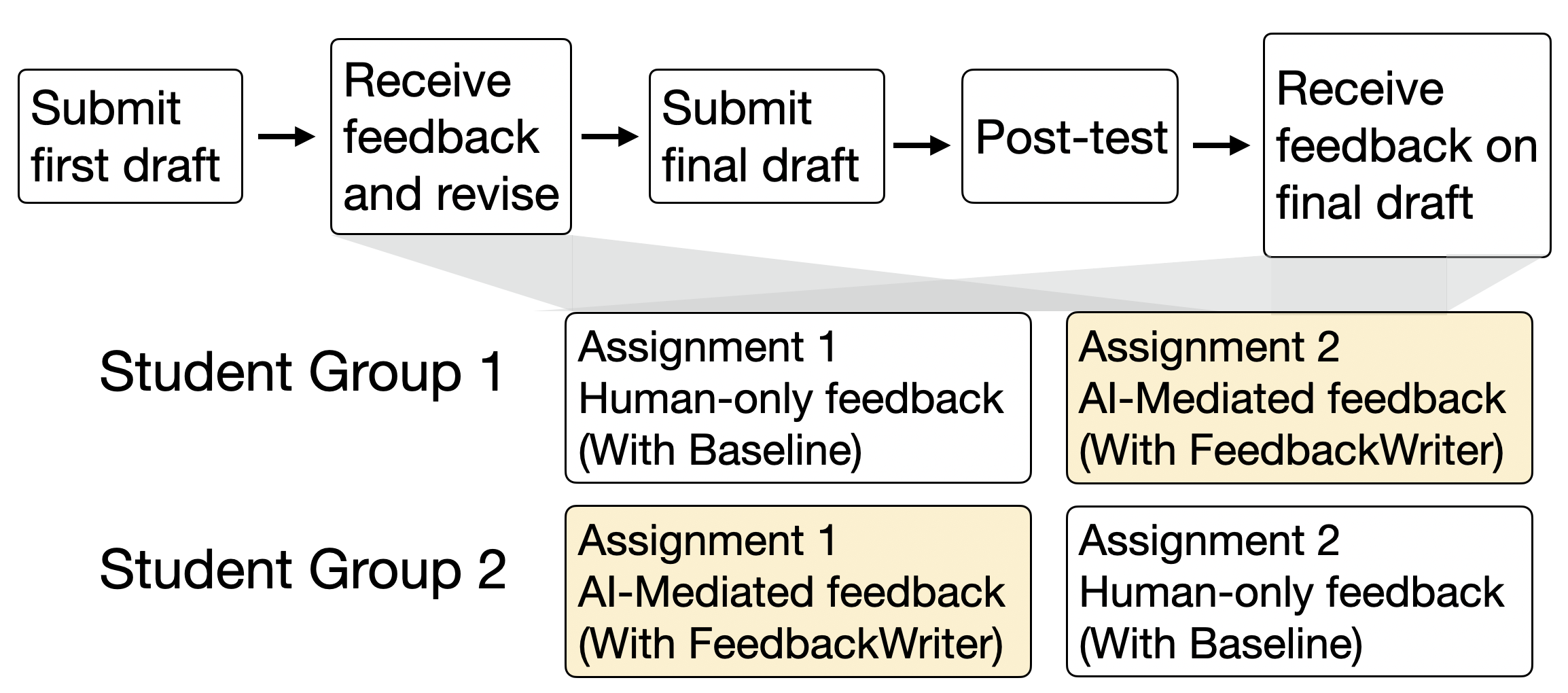}
    \caption{\highlight{Timeline for the study. For each assignment in ECON101, students submitted a first draft, received and incorporated TAs' feedback to form a final draft. The conditions were counter-balanced for the assignments.}}
    \Description{Diagram showing the study timeline and counterbalanced conditions. Timeline (left to right): submit a first draft, receive feedback and revise, submit a final draft, post-test, receive feedback on final draft. Below, two groups swap feedback conditions across assignments: Student Group 1 has Assignment 1 human-only feedback (with baseline) and Assignment 2 AI-mediated feedback (FeedbackWriter); Student Group 2 has Assignment 1 AI-mediated feedback (FeedbackWriter) and Assignment 2 human-only feedback (with baseline).}
    \label{fig:timeline}
\end{figure}

ECON101 is an introductory economics course with two writing prompts that apply course concepts to real events (e.g., select a news article from the past three months illustrating a market failure, identify the type, and propose a remedy). \highlight{We selected this course mainly because it is a college STEM gateway course with large enrollment, and it uses Writing-To-Learn assignments with detailed rubrics developed.} As shown in Figure \ref{fig:timeline}, for each prompt, students submit an initial draft, receive formative TA feedback, and submit a revised final draft for summative grading, yielding four submissions per student (two first drafts, two finals). The course enrolled 354 students in 11 sections; 11 TAs each supported one section and served as primary evaluators.

TAs were recruited by the lead instructor through a highly competitive process. All have achieved an A/A\textsuperscript{+} grade in the course, and completed a one-semester training course on providing feedback. To support quality and consistency, the lead instructor provided detailed rubrics, historic feedback from previous semesters as exemplars, and weekly meetings \highlight{for rubric calibration and }grade-norming sessions. 

\subsection{Ethics and Consent}
FeedbackWriter for ECON101 was iteratively refined in collaboration with two course instructors over two semesters; the lead instructor has taught the course 50+ times. 
The instructors reviewed AI-generated feedback in multiple rounds and confirmed that the feedback quality was reasonable before the deployment study. The design of FeedbackWriter made sure that students never saw raw AI output\highlight{---}all student-facing feedback needed to be explicitly inserted by the TAs. The instructors adopted FeedbackWriter as an instructional intervention, and the deployment was approved as educational-exempt by the university IRB and data stewards. This means that all TAs used the tool and all students received the resulting feedback. 
Students and TAs could opt out of the study, which means they would still follow the same practice, but the instructors would not share their data with the researchers. 

The course instructor announced the study during the first lecture and posted a written statement through Canvas \highlight{\footnote{\url{https://www.instructure.com/canvas}}}. It is emphasized that participation was voluntary and that opting out would not affect course standing or grades. 
No students or TAs of the class opted out. 
De-identified essays were processed via OpenAI’s API with no data retention.
The TAs received an incentive of \$100 to participate in the study throughout the semester. They also received an additional \$25 per hour for participating in the interviews.
The students did not receive any compensation.

\highlight{\subsection{Baseline Condition}}
\highlight{To compare the AI-mediated feedback with traditional human feedback, we designed a baseline system that is a visually matched, non-AI variant of the FeedbackWriter system. In the baseline condition, participants are provided with the same set of detailed rubrics and historic feedback as in the FeedbackWriter condition through a spreadsheet, which can be easily accessed through a link on the interface.}

\begin{figure*}
    \centering
    \includegraphics[width=\linewidth]{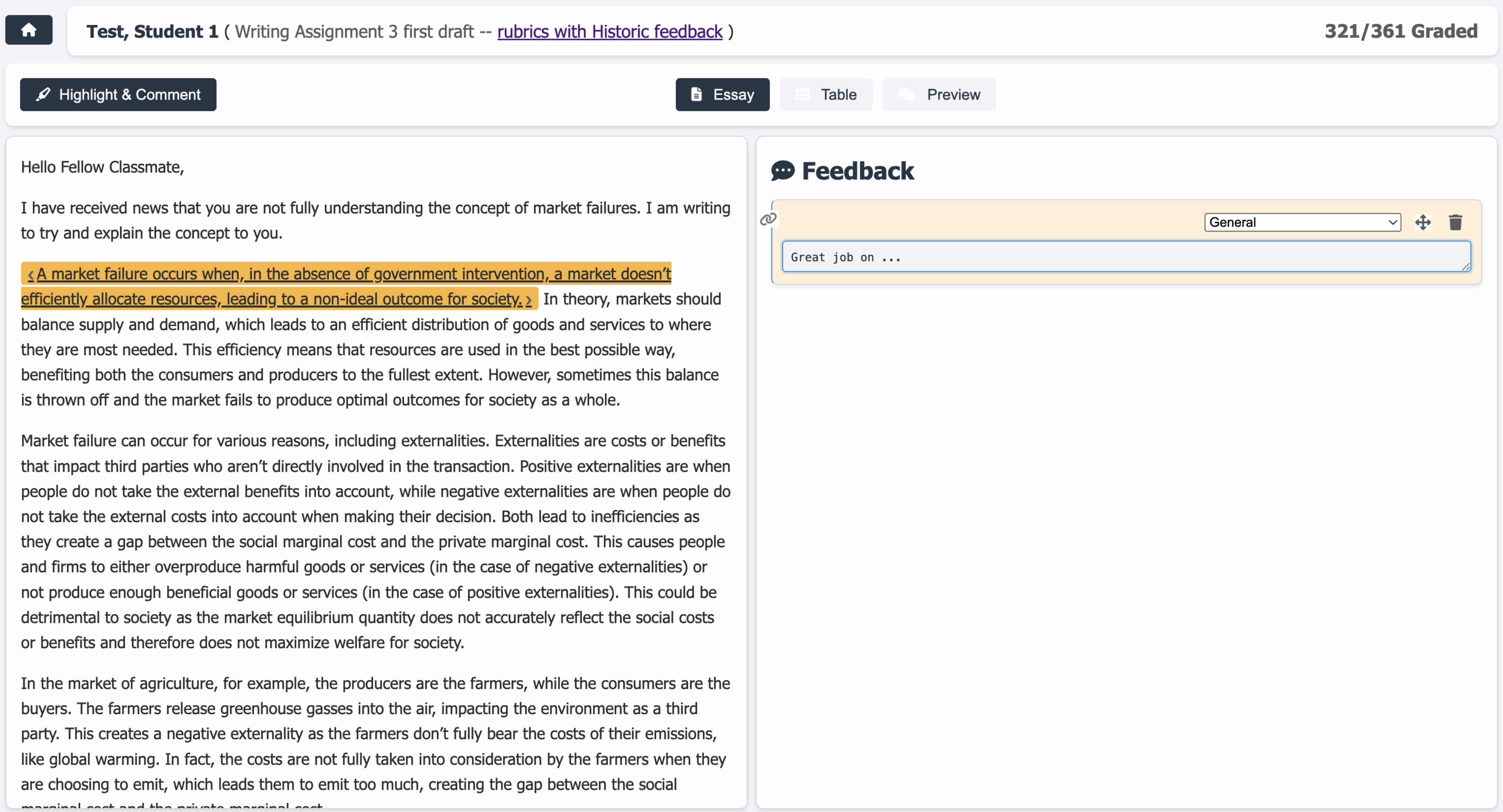}
    \caption{\highlight{The baseline condition is a visually matched, non-AI variant of the FeedbackWriter system. Users can create in-text comments and have easy access to the spreadsheet containing the rubrics and historic feedback, but there are no AI suggestions.}}
    \Description{Screenshot of the baseline interface. The main view is split into two panels: on the left, the student's essay text is displayed, with one sentence highlighted in orange. On the right, a "Feedback" panel shows only one comment box and a short draft feedback message beginning "Great job on ...".}
    \label{fig:baseline}
\end{figure*}

\highlight{The interface layout and interaction design are identical to those of FeedbackWriter, but no AI-generated suggestions are provided. The baseline condition represents a business-as-usual experience in feedback provision, consistent with the design of existing systems such as Canvas Speedgrader. The essay and feedback panel are displayed side by side (Figure~\ref{fig:baseline}). Users can create in-text comments by selecting text within a student essay, which automatically creates a feedback box anchored to the selected text for feedback. This setup mirrors common practices in current feedback provision workflows, as identified in our formative study.
}

\subsection{Study Design and Procedure}
The 11 TAs and their corresponding sections were randomly assigned to a condition for Writing Assignment 1: one group used the full FeedbackWriter system and the other used the non-AI variant (baseline). For Writing Assignment 2, we flipped the conditions so each TA experienced both. Students remained in their original sections for the semester. 

Prior to the study, students self-selected into discussion sessions only based on meeting time and seat availability. Pre-study exam scores show no significant difference in prior knowledge between conditions ($p=0.69$).

Students followed the standard course workflow. They submitted first drafts on Canvas, which were automatically imported into the FeedbackWriter system, where TAs provided formative feedback using their assigned conditions. The scores and comments generated in FeedbackWriter were posted back to Canvas via the Canvas API under each TA's credentials. 
Students then revised and submitted final drafts.
\highlight{On the day after the final draft is due,} students \highlight{received a post-test offered as an optional extra-credit quiz} to measure learning from the revision process. The quiz contains both multiple-choice and open-ended questions on knowledge components covered in the writing assignment. \highlight{In total, 262 students completed the post-test for Writing Assignment 1 (FeedbackWriter: 140/196; baseline: 122/157), and 305 students completed it for Writing Assignment 2 (FeedbackWriter: 133/159; baseline: 172/190). A proportion z-test showed no significant difference in participation rates between conditions.} TAs then followed the same process to grade the final draft under the same conditions. After each grading period, TAs can sign up for optional interviews with the researchers.

\subsection{TA Interviews}
After grading and providing feedback on each assignment, we invited the TAs to participate in a semi-structured interview to understand how they used the system, their challenges, and perspectives in the process. All interview sessions were conducted through Zoom, and participants were compensated with a \$25 Gift Card for each session. We conducted 24 sessions in total.

\subsection{Data Analysis}
\highlight{To answer RQ1 and RQ3, we analyzed quantitative outcomes using multivariate regression models (details in Section \ref{section:findings}). Prior work on writing intervention and feedback on writings typically shows a small to intermediate effect size, ranging from $0.25$ to $0.46$ \cite{kaliisa2025does, huisman2019impact, lv2021effects}. Since ECON101 is a large postsecondary course with heterogeneous instruction and partial compliance, we targeted an effect size of Cohen's $d=0.30$. With $\alpha = 0.05$, and $1-\beta=80\%$, the minimum number of students needed is $350$, and our final dataset included $354$ students.} 

\highlight{Statistic models were fit using the lme4 package in R \cite{lme4}. For all models, we first examined multicollinearity among predictors using Variance Inflation Factors (VIF). All independent variables had VIF values below the threshold of 5 \cite{xu2025peeredu}. We then conducted post-hoc diagnostic checks on model assumptions. Specifically, we inspected residual-versus-fitted plots to assess homoscedasticity and linearity, and Q-Q plots of residuals and random intercepts for students and TAs to assess normality. These diagnostics did not indicate violations of model assumptions. To answer RQ2, we compared the essay quality in the FeedbackWriter condition to the baseline condition. To answer RQ4 and RQ5, we performed affinity diagramming on the interview transcripts. The following of this section describes the design of data collection and processing.}

\subsubsection{System logs of TA interactions} 
FeedbackWriter records all user actions, including opening/closing assignments, creating or deleting a new highlight and feedback, modifying highlights or judgments, and adopting AI or historical feedback. 
Specifically, we aggregated the logs for each essay since revision quality is measured at the essay level. We extracted the following variable for each essay:
\begin{enumerate}[nosep]
    \item \textbf{\# Flip judgment:} number of times TAs flipped a judgment (met~$\leftrightarrow$~not met). This indicates when TAs don't agree with the AI's judgment.
    \item \textbf{\# Add historic feedback:} number of historical feedback inserted. All historic feedback comments are constructive. There are no positive praise in the historical corpus.
    \item \textbf{\# Add AI constructive feedback:} number of AI-generated constructive comments inserted (shown only when a rubric is not met).
    \item \textbf{\# Add AI \highlight{positive} feedback:} number of AI-generated praise comments inserted (shown only when a rubric is met).
    \item \textbf{\# Additional feedback:} number of extra comment boxes added beyond rubric-anchored comments.
\end{enumerate}

Timestamps on each action also allow us to estimate per-essay grading time as the elapsed time between the first and last action on that essay.

\subsubsection{Student Essay Quality Analysis.}
\label{sec:essayanalysis}
We used an LLM-based rubric scorer to evaluate both first and final drafts. For each rubric item, the model produced a satisfied/missing judgment (0 or 1) via the same rubric-based pipeline as FeedbackWriter; rubric-level scores were then aggregated into a total score using the TAs’ weighting scheme. We decided to use the AI-derived scores instead of the TA grades for the following four reasons:

\begin{enumerate}
    \item \textbf{TAs' first-draft grades were provisional.} TAs emphasized formative comments on first drafts, and those grades were hypothetical and not recorded.
    \item \textbf{High rater variance.} Despite multiple grade-norming sessions, TA scoring exhibited substantial between-rater variability.
    \item \textbf{Scalability.} A single uniform human rating for all essays was infeasible given class size.
    \item \textbf{Consistency.} Prior work reports higher internal consistency for AI-generated feedback than for human-authored feedback \cite{dai2023can}; similarly, from our data, AI rubric-level scores were more internally consistent than TA scores (Section~\ref{section:AIvsTA}). Moreover, system logs show that 88.7\% of rubric-level AI judgments matched TAs’ final judgments (Section~\ref{section:log}), providing convergent validity at the decision level. 
\end{enumerate}

\highlight{We used GPT-4o in the scorer because it was the best-performing model on benchmark datasets when this work was conducted. 
A potential concern with LLM-based scoring is generation–evaluation bias---a model may systematically favor text that it produced. Prior work suggests this bias is most pronounced when the same model both generates and evaluates AI-written text \cite{panickssery2404llm}. In our setting, however, the essays are written by students, and the AI feedback did not contain the expected solutions, but could only influence revisions indirectly.}
\highlight{To further assess whether the AI-derived scores faithfully reflect the essay quality, we recruited a human expert to manually evaluate 60 randomly sampled essays, with 30 from each assignment. The expert annotated each of the 35 rubrics per essay as satisfied/missing (0 or1), yielding 2,100 judgements in total. The expert spent more than 13 hours completing the task. Treating the expert ratings as the gold standard, the LLM-based rubric scorer (with GPT-4o) achieved a classification accuracy of 84.6\% and a recall of 85.6\%. We also compared against other high-performing models (GPT-5 and Gemini-3-Pro), which showed comparable accuracy and recall scores. GPT-5 achieved 85.8\% accuracy, and 87.7\% recall, and Gemini-3-Pro achieved 85.7\% accuracy and 87.9\% recall. 
These results suggest that the LLM-based rubric scorer powered by GPT-4o aligns reasonably well with expert judgments, and is a representative choice for evaluating essay quality in our study.}

\subsubsection{Quantitative Feedback Quality Analysis.}
\textbf{Evaluation Metrics:}
We extracted the final feedback messages each essay received after TAs' insertion and modification. Following prior frameworks on evaluating feedback quality, we devised desirable feedback properties and used an LLM-based pipeline to automatically tag them. First, following Patchan and Schunn's framework
\cite{Patchan_Schunn_Correnti_2016}, we categorized each feedback unit as Summary, Praise, Problem, and Solution (see Appendix \ref{appendix:categorization} for definitions and examples). As the framework suggested, \cite{Patchan_Schunn_Correnti_2016},feedback types influences the feedback effectiveness, e.g., Problem and Solution strongly predict student revision.

Second, we considered both the quality of tutoring efforts behind the feedback and the quality of the written content in the feedback. Drawing on the INSPIRE model \cite{Lepper_Woolverton_2002} and previous work on evaluating tutoring and feedback quality \cite{Kakarla_Thomas_Lin_Gupta_Koedinger_2024, Steiss_Tate_Graham_Cruz_Hebert_Wang_Moon_Tseng_Warschauer_Olson_2024}, we designed 4 metrics: content accuracy, promotion of independent learning, actionability, and tone and supportiveness (see Appendix \ref{appendix:feedback_rubric} for the detailed metric rubric and examples). 
Since independent learning and actionability presuppose diagnostic or prescriptive content, we applied these metrics only to Problem or Solution feedback units. We also excluded feedback units focused solely on prose mechanics (e.g. word count, citations, and conciseness) to concentrate on substantive, knowledge-intensive issues.

\textbf{Operationalization:}
Following prior automated analyses of feedback \cite{Patchan_Schunn_Correnti_2016}, we first segmented each feedback message into idea units, which are minimal units that address a single issue. 
Due to the large scale of the feedback dataset, we created an LLM-based evaluation pipeline using GPT-4.1 to (1) split feedback into idea units and map units to rubric item(s); (2) classify the feedback type; (3) assign binary ratings (0 or 1) for each quality metric using few-shot prompting. To evaluate the reliability of the pipeline, two researchers independently labeled 100 randomly sampled feedback (blinded to condition) for each step and resolved disagreements through discussion to produce the ground-truth labels. \highlight{We then evaluated the pipeline by computing Cohen’s Kappa between pipeline outputs and ground truth \cite{cohen1960coefficient}.}
The LLM-human agreement was high for all steps: For Step~1 (segmentation+linking), overall accuracy was $0.94$. For Step~2 (type classification), accuracy was $0.969$. For Step~3 (quality metrics): \emph{content accuracy} was $0.987$ ($\kappa=0.850$), \emph{promotion of independent learning} was $0.961$ ($\kappa=0.908$), \emph{actionability} was $0.974$ ($\kappa=0.901$), and \emph{tone and supportiveness} was $0.948$ ($\kappa=0.789$).

For each feedback segment, it has the following attributes:
\begin{enumerate} [nosep]
    \item \textbf{Feedback Type:} Summary, Praise, Problem, Solution (the types are not mutually exclusive). Each of these is a binary variable. 
    \item \textbf{Content Accuracy:} binary variable indicating whether the content is accurate or not.
    \item \textbf{Tone and Supportiveness:}  binary variable indicating whether the tone is supportive.
    \item \textbf{Actionability:} this only applies to the Problem/Solution type, a binary variable indicating whether the feedback is actionable.
    \item \textbf{Promotion of Independent Learning:} this only applies to the Problem/Solution type, a binary variable indicating whether the correct answer is revealed, and whether the student receives guidance.
\end{enumerate}

\subsubsection{Qualitative Analysis on Feedback Messages}
\highlight{\textbf{Comparison of the final feedback between conditions: }To better understand how the final feedback content differs between the FeedbackWriter condition (AI-mediated) and the baseline (human-only), two experts analyzed 50 pairs of feedback. Each pair contains feedback from both conditions targeting the same rubric item. We excluded AI-mediated feedback that directly reused historical feedback without edits, and randomly sampled 50 feedback pairs. We then conducted affinity diagramming \cite{lucero2015using} to identify differences in themes and patterns.}

\textbf{TA Modifications on AI Feedback: }
To understand how TAs use the AI feedback suggestions, we analyzed the modifications to feedback suggestions in the FeedbackWriter condition. For each instance, we extracted the original AI-feedback and the final TA-edited feedback to form paired examples, and performed qualitative content analysis of the edits to characterize how TAs revised the suggestions.

\subsubsection{Interview Data Analysis.} We transcribed the interviews and performed \highlight{affinity diagramming \cite{lucero2015using}. Two researchers jointly open-coded 2 sessions to generate initial codes, and incorporated feedback from co-authors. They then independently coded 3 additional sessions and met to reconcile differences and refine the codebook. Discrepancies were resolved through discussing rationales, revising code definitions, and jointly deciding the most appropriate code for each segment. Next, the two researchers coded the remaining 19 studies separately, and collaboratively performed affinity diagramming to iteratively group codes by semantic similarity, labeled clusters, and iteratively merged and split groups. This process produced the salient themes reported in RQ4 and RQ5.}

\section{Findings} \label{section:findings}
We present findings in response to each of the research questions. Here is a clarification \highlight{of the }variables in the statistical models:
\begin{enumerate} [nosep]
    \item \textbf{Final Draft Quality:} this is a score generated by the LLM-based rubric scorer, as introduced in Section~\ref{sec:essayanalysis}. The percentage score has a mean of 68\%, max of 100\%, and min of 17\%. 
    \item \textbf{First Draft Quality:} this is a score generated by the same LLM-based rubric scorer. This score is a proxy for the original draft quality. The percentage score has a mean of 48\%, max of 87\%, and min of 17\%. 
    \item \textbf{TA Score on First Draft (Motivational Factor):} this is a score assigned by the TAs on students' first drafts. This score is provisional and not recorded. However, students see this score before the revision. We treated this as a motivation factor, e.g., for students who received lower TA scores, they might feel more motivated to revise.
    \item \textbf{Post-test Score:} Student scores on the post-test. The post-test scores are manually graded blind to condition. The score is a proxy for student learning from the revision process. The percentage score has a mean of 58\%, max of 96\%, min of 21\%. 
    \item \textbf{Condition:} FeedbackWriter or Baseline.
    \item \textbf{Assignment-ID, Student-ID, TA-ID:} Assignment ID has two values A1 or A2, Student-IDs and TA-IDs are also recorded.
\end{enumerate}

Before addressing the research questions, we note two contextual results. First, a randomization check confirmed that students from the two conditions received comparable scores on two proctored in-person exams before the study. This indicates that there is no systematic difference between the two conditions prior to the study. 
Second, TAs spent comparable time grading in both conditions; despite assistance from FeedbackWriter, grading time did not decrease.
Section~\ref{sec:designsuggestion} outlines UI improvements that could streamline this workflow and reduce time in the future. 
These contextual results should inform the interpretation of the main findings that follow.

\subsection{RQ1: How does AI-mediated feedback affect students' revision and learning compared with human-only feedback? }

\subsubsection{Students who received AI-mediated feedback authored in FeedbackWriter had higher quality revised drafts} 
We built a mixed-effects linear regression model, with the Final Draft Quality as the dependent variable. Fixed effects include First Draft Quality and Condition. 
We added the TAs' score on the first draft (our “Motivational Factor”) and Assignment ID as covariates to adjust for motivation and assignment-specific difficulty.
Considering that different students may have different abilities, and different TAs may have different standards, we included random intercepts for each student and each TA. We used the lme4 R package \cite{bates2007lme4} to build the model, and the formula is shown below. 

\vspace{-0.7pc}
\begin{equation}\label{eq:1}
\begin{split}
 FinalDraftQuality = FirstDraftQuality + TAScore+ \\factor(AssignmentID)+ (1|StudentID)+ (1|TAID)
 \end{split}
\end{equation}

As shown in Table \ref{table:student_score}, we found that First Draft Quality strongly predicts the Final Draft Quality ($b=0.56,~t=10.99,~p<0.001^{***}$), whereas the Motivational Factor does not have an effect. 
Further, there is a significant effect of the Condition, which suggests that when a student is in the AI condition (FeedbackWriter), they are more likely to have higher quality revised drafts ($b=0.05,~t=5.02,~p<0.001^{***}$). Controlling for covariates and random effects, the coefficient suggests that for the same student, receiving AI-mediated feedback is associated with a 5\% higher final-revision quality than business-as-usual human-only feedback, which is roughly the gain from addressing two additional rubric items between the first and final drafts. The effect size is Cohen's d = 0.5, roughly equilavent to moving a student from the 50th to the 70th percentile. 

\begin{table}[t]
\caption{Results from the mixed-effects linear regression model on the students’ final draft quality}
\begin{tabular}{lcccc} \hline
Independent Variable                                                        & Estimate & t value & p                \\ \hline
(Intercept)                                                                 & 0.37  & 12.31 & $<0.001^{***}$\\
Condition (AI)                                                        & 0.05     & 5.02  & $<0.001^{***}$ \\
\begin{tabular}[c]{@{}c@{}}Assignment (Assignment 2)\end{tabular} & 4.87      & 11.07  & $<0.001^{***}$ \\
TA Score\highlight{---}Motivational Factor                                                   & 0.07     & 1.35    & $0.178$            \\
First Draft Quality                                                  & 0.56    & 10.99   & $<0.001^{***}$ \\ \hline
\end{tabular}
\label{table:student_score}
\end{table}

\begin{figure}
    \centering
    \includegraphics[width=\linewidth]{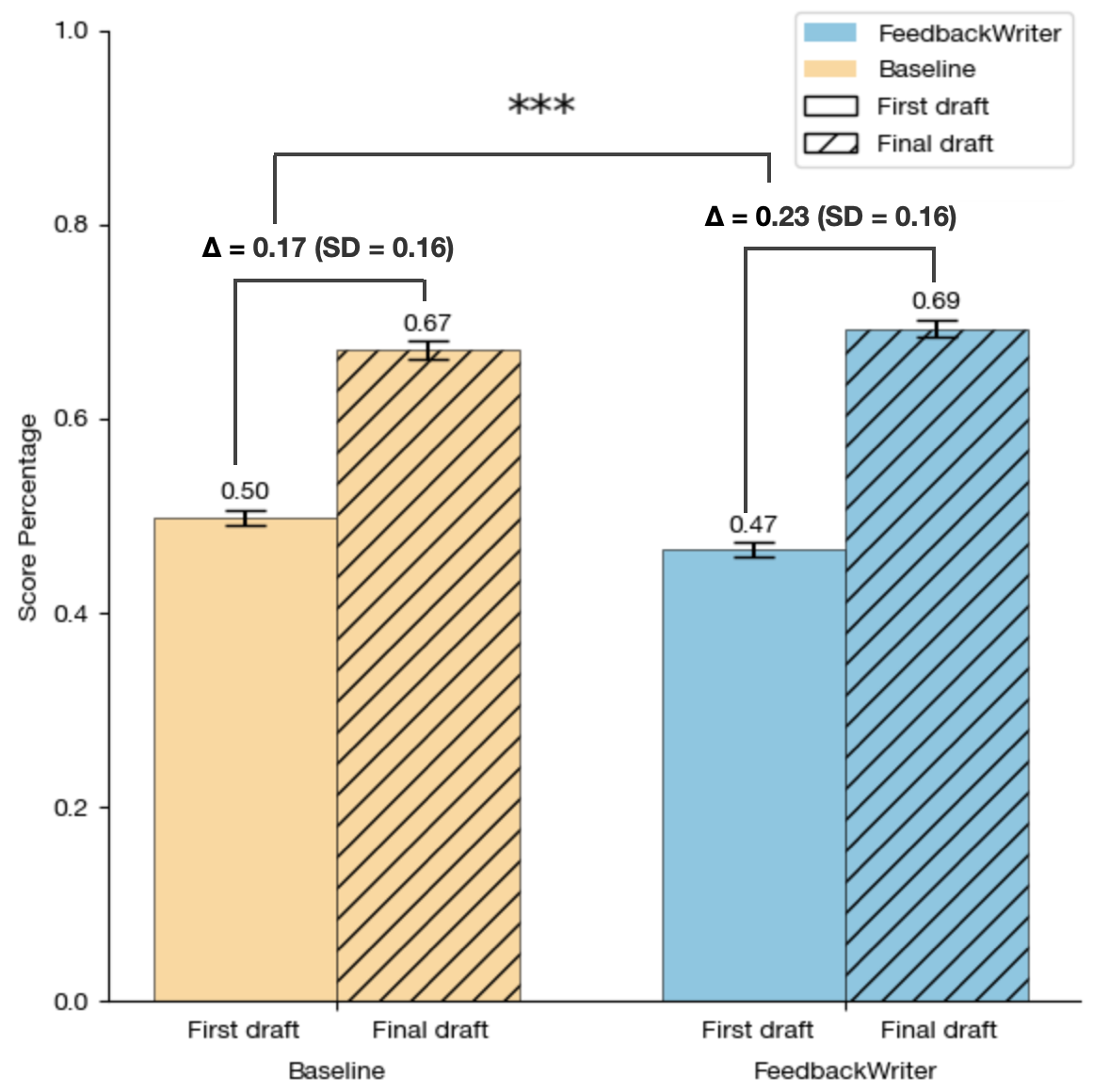}
    \caption{Draft quality by conditions. Students receiving AI-mediated feedback produced higher quality revisions, with significantly higher improvements, as measured by score difference from first to the final drafts ($p<0.001^{***}$).}
    \Description{Bar chart comparing average score percentage for first vs. final drafts in two conditions: FeedbackWriter (left) and Baseline (right). The y-axis is Score Percentage (0 to 1). Each condition has two bars with small error bars; the final-draft bars are hatched. In FeedbackWriter, the mean increases from 0.47 (first draft) to 0.69 (final draft), with an average improvement of 0.23 (SD = 0.16). In Baseline, the mean increases from 0.50 to 0.67, with an average improvement of 0.17 (SD = 0.16). A bracket across the two conditions with "***" indicates a statistically significant difference in improvement between conditions. }
    \label{fig:histogram_score}
\end{figure}

Descriptive statistics for score improvement from first to final drafts are displayed in Figure~\ref{fig:histogram_score}. In the FeedbackWriter condition, the mean score increases from $0.47~(SD=0.13)$ to $0.69~(SD=0.17)$, with an average improvement of $0.23~(SD = 0.16)$. In the baseline condition, the mean increases from $0.50~(SD=0.14)$ to $0.67~(SD=0.17)$, with an average improvement of $0.17~(SD = 0.16)$.

\subsubsection{Students receiving AI-mediated feedback from FeedbackWriter and human-only feedback had similar learning outcomes.}

We fit a second mixed-effects linear regression model, with the Post-test Score as the dependent variable.
Because students had submitted their final drafts before taking the post-test (but had not yet received grades), we included Final Draft Quality as an independent variable to capture variance in learning attributable to revision quality. Other independent variables include First Draft Quality, Condition, TAs' score on the first draft (our “Motivational Factor”) and Assignment ID.
We again included random intercepts for each student and each TA
. The formula is shown below. 

\vspace{-0.7pc}
\begin{equation}\label{eq:1}
\begin{split}
 PostTestScore = FinalDraftQuality+ FirstDraftQuality + \\TAScore+ factor(AssignmentID)+ (1|StudentID)+ (1|TAID)
 \end{split}
\end{equation}

\begin{figure}
    \centering
    \includegraphics[width=0.65\linewidth]{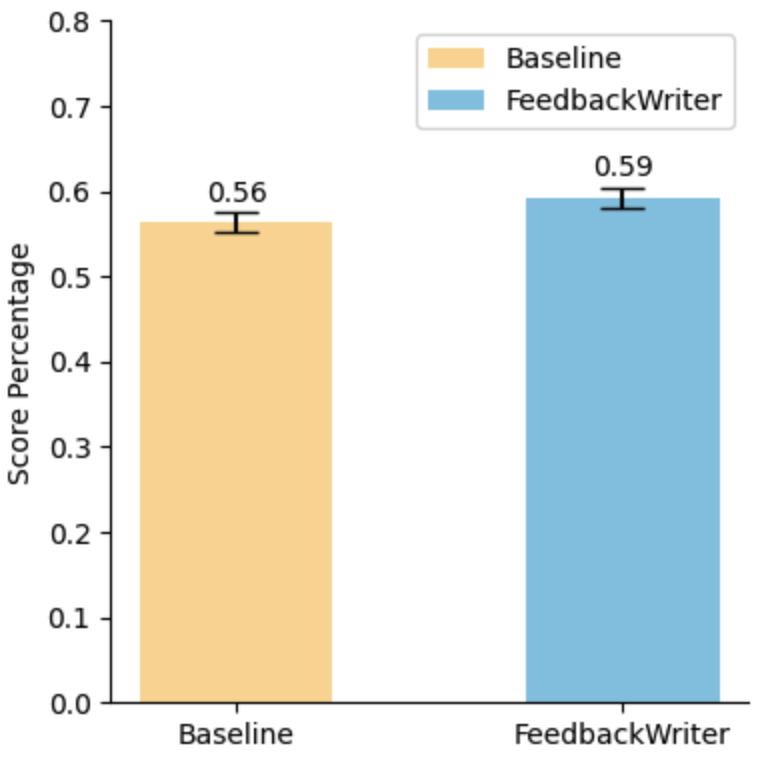}
    \caption{Students' post-test score percentage by condition. There was no significant difference between conditions.}
    \Description{Bar chart comparing mean score percentage for two conditions. The x-axis has FeedbackWriter and Baseline; the y-axis ranges from 0.0 to 0.8 (score percentage). FeedbackWriter has a mean score of 0.59, and Baseline has a mean score of 0.56. Both bars include small error bars, and FeedbackWriter is slightly higher than Baseline, but the difference is not significant.}
    \label{fig:histogram_posttest}
\end{figure}

As shown in Table \ref{table:posttest_score}, we found that being in the two different conditions did not affect students' post-test scores ($b=-0.007$, $t=-1.02$, $p=0.31$). Mean post-test scores were $0.59~(SD=0.19)$ for FeedbackWriter and $0.56~(SD=0.18)$ for Baseline.
Thus, although students in the FeedbackWriter condition produced higher-quality revisions, assignment to FeedbackWriter did not translate into additional post-test gains.
At the same time, final draft quality was a significant positive predictor of post-test performance ($b=0.09$, $t=3.60$, $p=<0.001^{***}$), indicating alignment between the post-test and the knowledge practiced in the writing task.
A plausible explanation for the null condition effect is limited practice dosage per knowledge component; for example, a rubric item such as "proposing a policy to address market failure (e.g., a tax or subsidy)" affords essentially a single practice opportunity during revision. Future designs should increase dosage by offering multiple, varied practice opportunities per rubric element. Descriptive statistics for the post-test scores are displayed in Figure~\ref{fig:histogram_posttest}. 

\begin{table}[t]
\caption{Results from the second mixed-effects linear regression model on the students’ post-test performance}
\begin{tabular}{lccc} \hline
Independent Variable                                                        & Estimate & t value & p                \\ \hline
(Intercept)                                                                 & 0.58  & 17.62 & $<0.001^{***}$\\
Condition (AI)                                                           & -0.007  & -1.02 & $0.31$ \\
\begin{tabular}[c]{@{}c@{}}Assignment (Assignment 2)\end{tabular} & -0.35      & -43.90  & $<0.001^{***}$ \\
TA Score\highlight{---}Motivational Factor                                                   & 0.02     & 0.61    & $0.54$            \\
First Draft Quality                                                   & 0.05     & 1.50    & $0.13$            \\
Final Draft Quality                                                  & 0.09    & 3.60   & $<0.001^{***}$ \\ \hline
\end{tabular}
\label{table:posttest_score}
\end{table}



\subsection{RQ2: How does feedback quality differ between the FeedbackWriter (AI-mediated) and baseline (human-only) conditions?}

\subsubsection{When grading using FeedbackWriter (AI-mediated), TAs provide more feedback messages.}
When TAs used FeedbackWriter, they provided more pieces of feedback. For the first draft, TAs using FeedbackWriter provided 10.53 feedback messages on average for Writing Assignment 1 (n=196), and 15.13 feedback messages for Writing Assignment 2 (n=158). In contrast, TAs in the baseline provided 6.6 feedback messages on average for Writing Assignment 1, and 7.5 feedback messages for Writing Assignment 2 (n=190).

\subsubsection{When grading using FeedbackWriter (AI-mediated), TAs exhibit higher inter-rater consistency} \label{section:AIvsTA}

\begin{figure*}
    \centering
    \includegraphics[width=\linewidth]{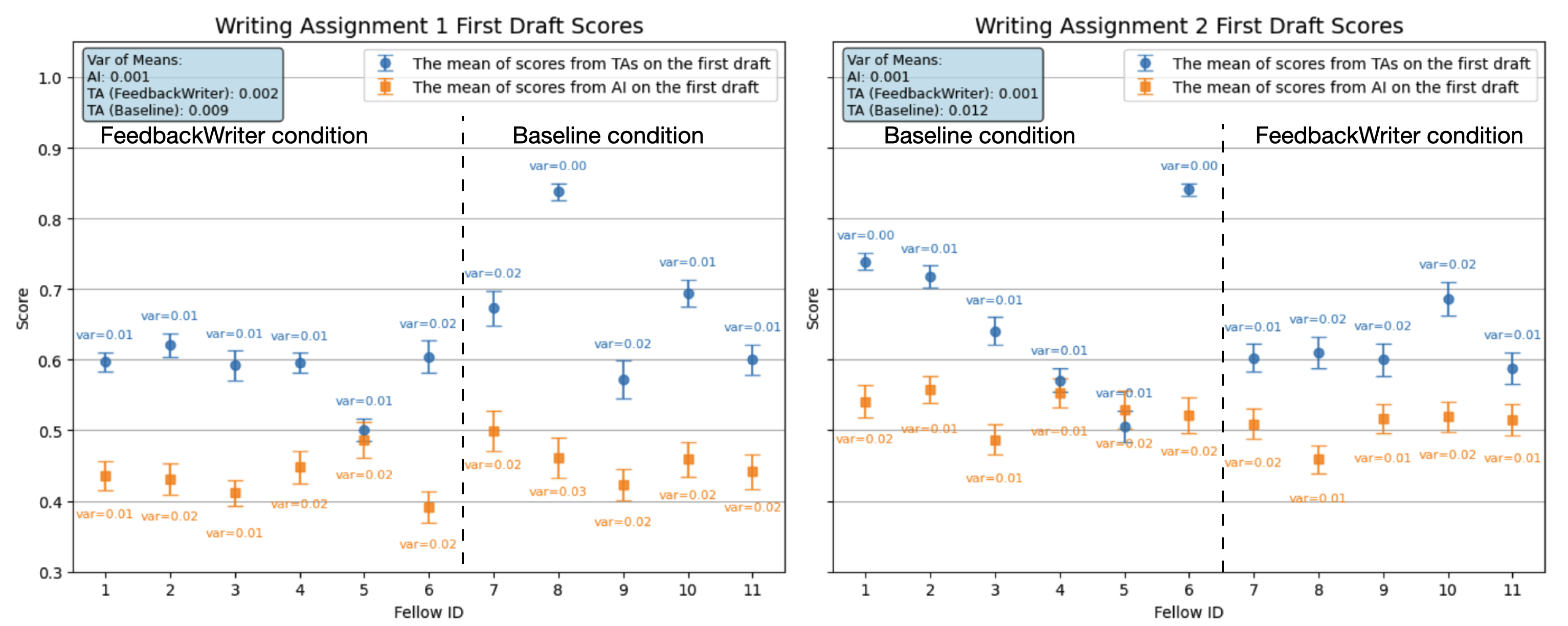}
    \caption{TAs in the baseline had higher inter-rater variance, whereas TAs in FeedbackWriter were more consistent. AI scores had lower variances compared with TAs in the baseline. This is one of the reasons that we did not use the TA scores to indicate essay quality.}
    \Description{Two-panel scatter plot with error bars comparing TAs' first-draft writing scores versus the AI-derived scores across 11 fellows for Writing Assignment 1 (left) and Writing Assignment 2 (right). In each panel, the x-axis is Fellow ID (1-11) and the y-axis is Score (ranging 0.3 to 1.0). Each plot include blue circles and orange squares showing the mean TA score per fellow and the corresponding mean AI score on the same submissions, with error bars. Small text labels next to points report per-fellow variance values. Across both assignments and conditions, TA mean scores are consistently higher than AI mean scores for the same fellow. Insets summarize the variance of the mean values: AI = 0.001 in both panels; TA variance is smaller in FeedbackWriter (0.002 in Assignment 1; 0.001 in Assignment 2) than in Baseline (0.009 in Assignment 1; 0.012 in Assignment 2).}
    \label{fig:variance}
\end{figure*}

As shown in Figure \ref{fig:variance}, TAs in the baseline condition had higher inter-rater variance, whereas TAs in the FeedbackWriter condition were more consistent in terms of the scores they assigned. Moreover, AI-generated scores had lower variances in comparison to the TAs in the baseline condition. This is one of the reasons that we did not use the TA (human-only) scores to indicate essay quality.

\subsubsection{When grading using FeedbackWriter (AI-mediated), TAs provide longer feedback with higher coverage of the rubric items.}
\begin{figure}
    \centering
    \includegraphics[width=\linewidth]{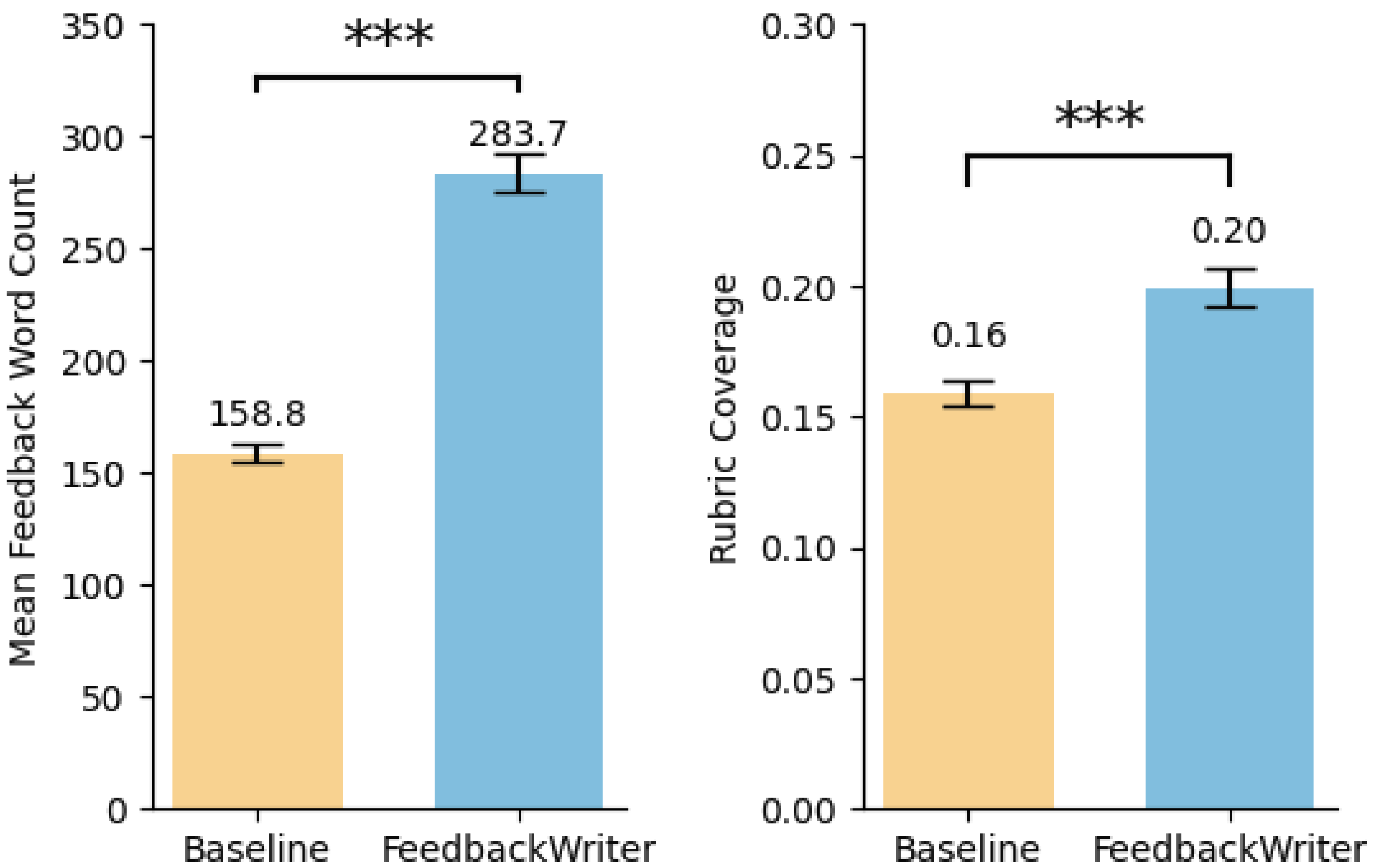}
    \caption{AI-mediated feedback from FeedbackWriter has significant higher word counts (left) and coverage of the rubrics (right) on average than the baseline ($p<0.001^{***}$).}
    \label{fig:feedback_count}
    \Description{Two-panel bar chart. The bar chart on the left compares mean feedback word count for two conditions (FeedbackWriter in blue, Baseline in orange). The x-axis lists the two conditions; the y-axis shows Mean Feedback Word Count (ranging 0 to 350). Mean values are labeled above bars: FeedbackWriter = 283.7 words, Baseline = 158.8 words, each with error bars. A bracket with “***” above the two bars indicates a statistically significant difference, with FeedbackWriter higher. The bar chart on right compares rubric coverage for two conditions (FeedbackWriter in blue, Baseline in orange). The x-axis lists the two conditions; the y-axis shows Rubric Coverage (ranging 0 to 0.35). Mean values are labeled above bars: FeedbackWriter 0.20, Baseline 0.16, each with error bars. A bracket with “***” above the two bars indicates a statistically significant difference, with FeedbackWriter higher.
    }
\end{figure}

When controlled for Assignment-ID, and adding random intercepts for Student-ID and TA-ID, feedback resulting from the FeedbackWriter condition show significantly higher coverage on the rubrics ($M = 0.195,~ SD=0.14$ vs.\ $M=0.165,~ SD=0.09$, $p < 0.0001^{***}$) and contain significantly more words ($M = 286,~ SD=161$ vs.\ $M=157,~SD=77$, $p < 0.0001^{***}$), as shown in Figure \ref{fig:feedback_count}.

\begin{figure}
    \centering
    \includegraphics[width=0.95\linewidth]{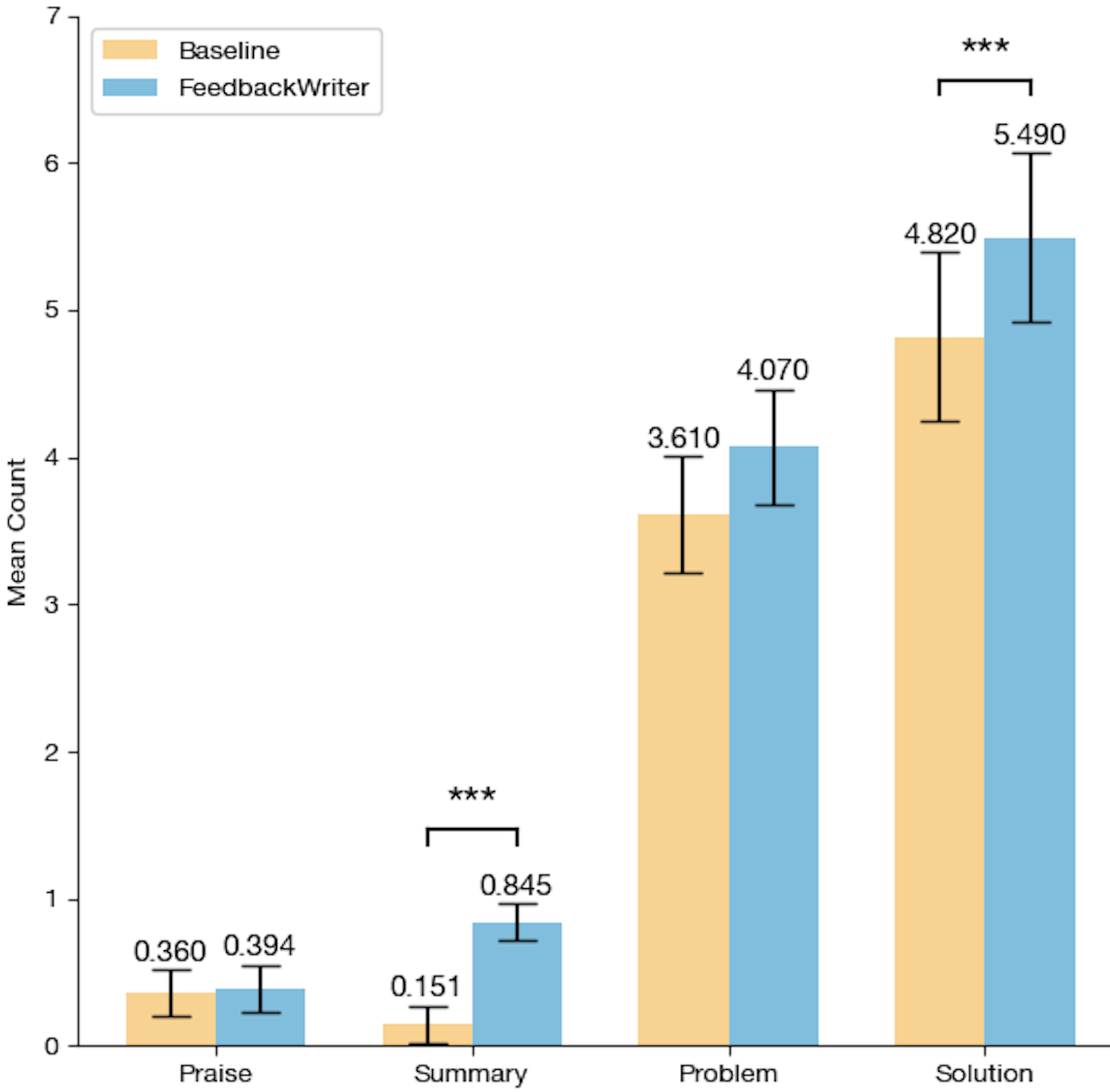}
    \caption{Category distributions of the feedback by condition. The y-axis is the average number of each type per essay. AI-mediated feedback from FeedbackWriter condition had significantly more Summary and Solutions.}
    \Description{Four groups of bar chart showing the mean count (with error bars) of four feedback categories for two conditions (FeedbackWriter in blue, Baseline in orange). The x-axis lists the four categories: Praise, Summary, Problem, and Solution. The y-axis shows the mean count (ranging 0 to 7). Mean values labeled above bars: For praise, FeedbackWriter 0.394, Baseline 0.360; For summary, FeedbackWriter 0.845, Baseline 0.151; For problem, FeedbackWriter 4.070, Baseline 3.610; For solution, FeedbackWriter 5.490, Baseline 4.820. Brackets with “***” indicate statistically significant differences for Summary and Solution, with FeedbackWriter higher in both.}
    \label{fig:feedback_ratings_category}
\end{figure}

For each feedback message, we tagged its type into Summary, Praise, Problem, and Solution (see Appendix \ref{appendix:categorization} for the detailed metric rubric and examples). 
As shown in Figure \ref{fig:feedback_ratings_category}, for each essay, the TAs using FeedbackWriter provided significantly more Summary ($M = 0.845,$ vs.\ $0.151$, $p < 0.0001^{***}$) and significantly more Solution ($M = 5.49$ vs.\ $4.82$, $p = 0.0003^{**}$), while the differences in Praise ($M = 0.394$ vs.\ $0.360$, $p = 0.394$), and Problem ($M = 4.07$ vs.\ $3.61$, $p = 0.022$) are not significant.

\subsubsection{Feedback messages resulted from the FeedbackWriter condition exhibit more desirable properties for effective feedback.}

\begin{table*}
\caption{\highlight{Example pairs of feedback from FeedbackWriter (AI-mediated) and baseline (human-only) condition with analysis and themes. In the ``AI-mediated feedback'' column, the bolded texts visualize the original AI-generated suggestions adopted in the final feedback. In the ``Human feedback'' column, the bolded texts indicate the historic feedback adopted in the final feedback.}}
\begin{tabular}{|p{0.14\linewidth}|p{0.25\linewidth}|p{0.16\linewidth}|p{0.13\linewidth}|p{0.23\linewidth}|}
\hline
Student mistake                                                                                                                                             & AI-mediated feedback                                                                                                                                                                                                                                                                                                                                                                                                                                                                                                                  & Human feedback                                                                                                                                                                                                    & Themes                                                                                   & Analysis                                                                                                                                                                                                                                                                                                                                                 \\ \hline
The student mentioned ``substitutes'' but not ``substitutes in consumption'' as required
& \textbf{You've identified the role of substitutes, but how might you further explore the relationship between almond milk and its alternatives in consumer choices?} What kind of substitutes are they?
& As well I would mention the idea of what alternatives are in the market.
& Acknowledgement of progress
& The AI-mediated feedback acknowledges what the student already did before suggesting the next step. The hint contains questions, which clearly direct how the student could improve, without revealing the answer. \\ 
\hline

The student didn’t analyze how the solution changes the artificially scarce good to a public good.
& \textbf{What would happen if insulin were priced at its marginal cost? Consider how this might change its classification and the potential issues that could arise.}
& \textbf{Consider the solutions discussed in class. What would happen if the government required the good be sold at the efficient price?}
& More specific, personalized, and localized
& The AI-mediated feedback gives a specific and clearer direction for critical thinking and deeper analysis, tailors to the essay topic, and localizes the feedback to where the student discussed the insulin price. \\ 
\hline

The student mentioned tax, but did not discuss the efficient size
& \textbf{How can the efficient size of a tax be determined, and why is this important for addressing the negative externality?}
& \textbf{Remember to expand well on a thoughtful, well-reasoned solution to the negative externality you present in your article.}
& More tailored to progress
& When the student partially meets the rubric item, the AI-mediated feedback is more tailored to the progress, while the human-only feedback remains generic to the rubric item. \\ 
\hline

Mistake demand vs.\ quantity demanded
& \textbf{Can you clarify the difference between a change in demand and a change in quantity demanded?} Remember that stating a ``decrease in demand'' implies a demand shift, which does not embody the true economic impact of the tax. \textbf{How does a price increase from the tax specifically affect quantity demanded?} It may help to draw out the tax on the water market on a graph and mention the changes that you see to price and quantity.
& Water demand would not decrease as a result of the policy, but there would be a decrease in quantity demanded.
& More specific; not revealing the answer
& The AI-mediated feedback doesn't reveal the answer, but explains the mistake and suggests a concrete tool (graph) to approach the problem. The human-only feedback only points out the mistake without supporting the student to reflect, understand, or improve. \\ 
\hline

\end{tabular}
\label{table:ai-mediated-vs-human}
\end{table*}

\begin{figure}
    \centering
    \includegraphics[width=\linewidth]{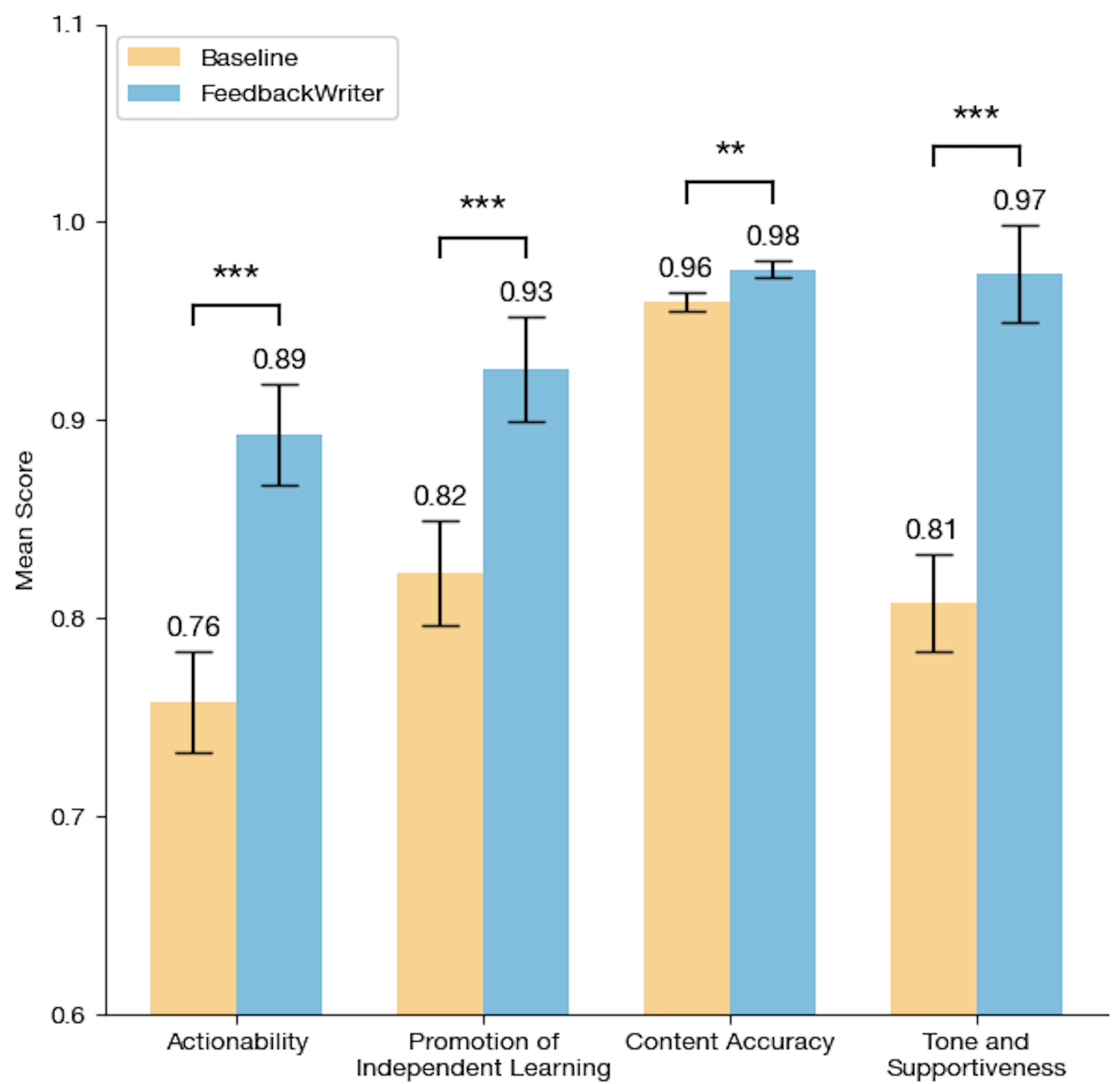}
    \caption{Mean scores on desirable properties of effective feedback by condition. The y-axis is the percentage of feedback comments that have this attribute. AI-midated feedback from FeedbackWriter were rated higher on all properties.}
    \Description{Four groups of bar chart showing the mean score (with error bars) for four feedback quality characteristics for two conditions (FeedbackWriter in blue, Baseline in orange). The x-axis lists Actionability, Promotion of Independent Learning, Content Accuracy, and Tone and Supportiveness. The y-axis shows the mean score (ranging 0.6 to 1). Mean values labeled above bars: For Actionability, FeedbackWriter 0.89, Baseline 0.76; For Promotion of Independent Learning, FeedbackWriter 0.93, Baseline 0.82; For Content Accuracy, FeedbackWriter 0.98, Baseline 0.96; For Tone and Supportiveness, FeedbackWriter 0.97, Baseline 0.81. Brackets above each pair indicate statistically significant differences: “***” for Actionability, Promotion of Independent Learning, and Tone and Supportiveness, and “**” for Content Accuracy, with FeedbackWriter higher than Baseline in all four characteristics.}
    \label{fig:feedback_ratings}
\end{figure}

We compared 3,548 feedback units from FeedbackWriter (Writing Assignment 1: n=1,894, Writing Assignemnt 2: n=1,654) condition with 2,794 feedback units from the baseline (Writing Assignment 1: n=1,226, Writing Assignemnt 2: n=1,568). As shown in Figure \ref{fig:feedback_ratings}, specifically, AI-mediated feedback was rated significantly higher in \textbf{actionability} ($M = 0.893$ vs.\ $0.758$, $p < 0.0001^{***}$), \textbf{promotion of independent learning} ($M = 0.926$ vs.\ $0.823$, $p < 0.0001^{***}$), and \textbf{tone and supportiveness} ($M = 0.974$ vs.\ $0.808$, $p < 0.0001^{***}$). \textbf{Content accuracy} showed less significant difference between AI and baseline conditions ($M = 0.976$ vs.\ $0.960$, $p=0.0006^{**}$), likely because the TAs' expertise was comparable across both conditions. 

\highlight{Our qualitative comparison of AI-mediated and human-only feedback shows similar patterns (Table \ref{table:ai-mediated-vs-human}). 
AI-mediated feedback was more closely tailored to each student’s essay content. For instance, an AI-mediated comment such as, ``What would happen if insulin were priced at its marginal cost?'' explicitly builds on the student’s example (insulin), whereas the baseline comment is more generic, saying, ``Consider the solutions discussed in class. What would happen if the government required the good be sold at the efficient price?''.
AI-mediated feedback also more often acknowledged students’ partial progress before suggesting next steps (e.g., ``You’ve mentioned the gap between social and private marginal costs, but what exactly are external marginal cost and social marginal cost?''), which is more supportive and actionable, as it surfaces why the existing content is not sufficient.}


\subsection{RQ3: How do TAs engage with FeedbackWriter's AI features while authoring feedback?}

\subsubsection{TAs actively and critically utilized the AI suggestions} \label{section:log}

In this section, we report TAs' usage patterns in the FeedbackWriter condition based on the logs. 
Among the 703 essays in the FeedbackWriter condition (n=386 for Writing Assignment 1; n=317 for Writing Assignment 2), TAs evaluated 24,853 rubric items \highlight{and wrote} 6,121 feedback comments.

TAs actively incorporated AI suggestions. Historic feedback suggestions were adopted 1,348 times (22.2\% of feedback), and AI feedback suggestions 3,141 times (51.3\% of feedback). \highlight{Per essay, }TAs adopted on average 3.41 AI constructive comments, 0.91 AI positive comments, and 1.92 historic comments (Table \ref{table:operations}), with more feedback on first drafts \highlight{to support revision}. 

TAs did not accept AI suggestions blindly. Instead, they edited AI judgments when needed. 
On average, TAs flipped 2.83 AI's judgments per essay. 
Overall, 21,523 judgments (88.7\%) were approved and 2,747 (11.3\%) were corrected. More judgment modifications occured in Writing Assignment 1 (details and analysis in Section \ref{section:judgment}), \highlight{reflecting instructors' view that Writing Assignment 2 has rubrics that are} more straightforward and better broken down.

TAs used highlight repositioning 11 times and regenerated feedback 230 times.
They also further modified 806 feedback comments after adopting AI or historic feedback. We detail how TAs used and modified the AI suggestions in Section \ref{section:how_ai_suggestions_are_used}.

\begin{table*}[htbp]
\caption{In the FeedbackWriter Condition, TAs actively engaged with AI. The table displays the average count of actions for an individual essay.}
\begin{tabular}{lrrrrrr}
\hline
   & \begin{tabular}[c]{@{}r@{}}Modify \\ judgments\end{tabular} & \begin{tabular}[c]{@{}r@{}}Add historic\\ feedback\end{tabular} & \begin{tabular}[c]{@{}r@{}}Add constructive\\ AI feedback\end{tabular} & \begin{tabular}[c]{@{}r@{}}Add positive\\ AI feedback\end{tabular} & \begin{tabular}[c]{@{}r@{}}Additional\\ feedback \end{tabular} & \begin{tabular}[c]{@{}r@{}}Total\\feedback\end{tabular} \\ \hline
\begin{tabular}[c]{@{}l@{}}Writing Assignment\\1 first draft (n=196)\end{tabular} & 4.74                                                        & 2.79                                                            & 3.49                                                                   & 0.14                                                               & 1.22                                                                                & 10.53          \\
\begin{tabular}[c]{@{}l@{}}Writing Assignment\\1 final draft (n=190)\end{tabular}  & 4.38                                                        & 0.07                                                            & 4.03                                                                   & 0.14                                                               & 0.55                                                                                & 5.26           \\
\begin{tabular}[c]{@{}l@{}}Writing Assignment\\2 first draft (n=158)\end{tabular} & 0.67                                                        & 4.91                                                            & 4.01                                                                   & 3.68                                                               & 2.28                                                                                & 15.13          \\

\begin{tabular}[c]{@{}l@{}}Writing Assignment\\2 final draft (n=159)\end{tabular}  & 0.77                                                        & 0.08                                                            & 1.99                                                                   & 0.01                                                               & 1.17                                                                                & 4.21           \\
All assignments                                                                   & 2.83                                                        & 1.92                                                            & 3.41                                                                   & 0.91                                                               & 1.27                                                                                & 8.71           \\ \hline
\end{tabular}
\label{table:operations}
\end{table*}

\subsubsection{When TAs used FeedbackWriter and adopted more AI-generated constructive feedback messages, students had higher quality revisions.}

In response to RQ1, we found that when students received AI-mediated feedback authored from FeedbackWriter, they had higher quality revisions. Here, we further explore how TAs' interaction with FeedbackWriter might have contributed to that effect. We fit a third mixed-effects linear regression model with a similar setup as the first model. The Final Draft Quality is the dependent variable, and the covariates include Condition, First Draft Quality, and Assignment-ID. In addition we added five covariates that indicate TAs' interactions in the system. We again added random intercepts for each student and TA.

As shown in Table~\ref{table:score_operations}, the results suggest that when the TAs take more AI-suggested constructive feedback which are targeted at student mistakes, the student had higher revision quality ($b=0.005$, $t=2.22$, $p=0.027^{*}$). The other interactions do not predict the performance.  

\begin{table}[htbp]
\caption{Regression analysis of AI suggestion usage on the students' final draft quality}
\begin{tabular}{lcccc}\hline
Independent Variable                    & Estimate & t value & p                \\ \hline
(Intercept)                             & 0.36    & 10.7    & $<0.001^{***}$ \\
Condition (AI)                          & 0.04   & 2.44    & $0.015^{*}$             \\
Assignment (Assignment 2)       & 0.13   & 9.53    & $<0.001^{***}$ \\
Add Historic Feedback       & -0.003  & -1.44   & 0.15             \\
Add AI Positive Feedback & -0.0005  & -0.23    & 0.81             \\
Add AI Constructive Feedback     & 0.005 & 2.22   & $0.027^{*}$ \\
Additional Feedback                  & 0.002  & 1.15    & 0.25    \\ \hline        
\end{tabular}
\label{table:score_operations}
\end{table}

\subsection{RQ4: When do TAs accept, edit, or dismiss AI suggestions, and what considerations drive these choices?} \label{section:how_ai_suggestions_are_used}
Aligned with the results from log analysis, participants generally found the AI suggestions accurate and helpful. \highlight{When mistakes occurred,} participants found them easy to verify and correct in FeedbackWriter. \highlight{Below}, we present interview findings on how TAs decided to adopt, correct, or ignore the AI suggestions for different AI features.

\subsubsection{AI feedback is well-worded, specific and personalized, while historic feedback is trustworthy.}

Participants appreciated having both AI and historic feedback side-by-side, noting it reduced the need to write new comments from scratch. Participants praised AI feedback for being well-phrased, concise, and tailored to the student’s specific progress, often \textit{``strik[ing] a good balance \dots leading the student to the correct answer without giving it (the answer) away''} (P2). Several TAs found AI feedback clearer and more specific than historic feedback, especially on nuanced rubrics, explaining why a missing element mattered (P5, P11). Historic feedback, by contrast, was valued for trustworthiness and breadth. Because it is pre-approved and reused, TAs felt confident that it aligned with the instructor's intent (P4). Additionally, its generic framing nudged students to scan the whole essay for recurring issues, which is especially preferable for cross-cutting concerns like terminology usage (P1).

\subsubsection{TAs make further modifications on the feedback suggestions}
Participants rarely used the feedback regenerate feature because they found the initial AI feedback typically ``good enough'' (P1, P5), so they preferred making small manual edits over requesting another suggestion. Here, we discuss the most prevalent manual modifications.

\textbf{Add clarifying hints and localize where to revise.} 
Since the AI feedback is instructed to avoid revealing answers, TAs added brief cues when they worried students might be unsure how to improve. For instance, P4 referenced specific lines from the assignment prompt when a key factor was missing in their analysis. P11 appended \textit{``Hint\highlight{---}it changes the type of good''} to the AI feedback to focus the direction of revision. When students seemed stuck, TAs sometimes prioritize clarity over indirectness and non-revealing (P3, P5, P11). Moreover, TAs added quotes or section titles from the essay to help students anchor where to revise (e.g., ``after the policy'' (P9), ``when you stated \dots'' (P5, P10). As P5 explained, \emph{“sometimes it's relevant for the student to know where in the essay that I'm talking about \dots mainly whenever students got something wrong.”}

\textbf{Clarify gaps in near-miss mistakes.}
Because AI feedback is constrained to avoid revealing answers and to stick closely to the rubrics, its suggestions can be too generic to be helpful in ``near-miss'' cases, e.g., when a student uses an imprecise variant of a term. In these situations, students might incorrectly believe they have satisfied the criterion, given the generic guiding question in AI feedback. TAs therefore inject targeted cues that explain why the current text is insufficient. For example, one student identified 2 goods as ``substitutes'', while the rubric requires the term ``substitutes in consumption''. P2 adopted the AI feedback and added a clarification cue, ``What kind of substitutes are they?''. Similarly, P1 added a hint to an AI feedback, ``you mention deadweight loss, but fail to really explain what it means.''

\textbf{Extend the feedback to cover multiple relevant rubrics.} 
When there were higher-level misunderstandings of the assignment requirements or core concepts (e.g., missing welfare analysis or mistaking the terms ``demand'' and ``quantity demanded''), TAs replaced several rubric-level feedback with one overarching comment (P4, P4, P11) and consolidated adjacent rubrics (e.g., demand/supply vs. quantity demanded/supplied). The aim was to redirect the student's attention to the underlying missing concept and its role in analysis, rather than to surface details (P1, P3, P5). TAs found this improvement especially important when students were \textit{``not on the right track''} (P3) or had omitted an entire concept (P1), where detail-oriented AI feedback would not help address the high-level misunderstandings.

\textbf{Shorten or simplify the feedback.}
For checklist-style criteria (e.g., ``list all four types of market failures''), TAs trimmed verbosity to keep feedback scannable. As P11 noted, students are busy and tend to skip long comments, so they would benefit more from guidance that is \textit{``short and to the point''}. To achieve this, many participants removed explanatory elements, like the importance of the missed point (P1-P5, P9, P11), acknowledgements on partially correct answers (P4, P5, P7, P11), and next-steps when commenting on the final drafts (P2, P3, P9). TAs also simplify the language using phrases and abbreviations, such as replacing the term ``marginal benefit'' with ``MB'' (P10, P11).

\subsubsection{TAs treat AI judgments as a complement, not a replacement, for their own judgments.} \label{section:judgment}
As all participants were experts in the domain knowledge, they found it easy to make their own judgments on rubrics. Moreover, all participants mentioned that they kept in mind that AI could make mistakes, so they made their own judgments before verifying the AI judgments. Some participants reported spending extra time verifying when AI made mistakes in judgments. For example, P8 noted, \textit{``I would read through that certain part highlighted ... and read around that ... to kind of understand why, AI is flagging it (differently).''}

However, participants appreciated having the AI judgments as their complement. They found the AI judgment mostly accurate, and helped catch mistakes they missed. As P5 noted, they found the time spent to understand the AI judgment \textit{``minimal compared to the time that it does save when it can catch a mistake that you missed.''} Moreover, all participants used the judgment to assign scores on student essays, which motivated them to correct the judgments.

\subsubsection{\highlight{Modifying AI highlights and regenerating feedback are rarely used because these actions do not directly modify final outputs.}}

\highlight{Participants rarely correct the AI highlights because they did not treat highlighting as a standalone goal, but as a means to locate evidence for making their judgments (P1, P2, P3, P5). Once they located the relevant text and decided whether a rubric item was satisfied, they saw little value in refining the highlight. In contrast, judgments were corrected because they were later used to assign scores. Similarly, participants viewed AI-generated feedback suggestions as intermediate drafts, which are only used to inform the final feedback. Because participants found it easy to improve manually, they seldom requested another AI feedback suggestion (P2, P4, P9). As P4 explained, \textit{``Sometimes I knew exactly what I needed the student to say. And ... it's just easier for me to type it in rather than regenerate something and assess.''} As a result, participants more frequently improved judgments and final feedback than the intermediate suggestions, such as highlights or AI-feedback.}

\subsection{RQ5: How do TAs perceive FeedbackWriter and the human-AI teaming approach in their feedback provision process?}\label{sec:rq5}
Participants preferred having FeedbackWriter and the AI assistance when providing feedback. They appreciated the quality and convenience of the AI feedback compared with writing and highlighting on their own. P1 said, \textit{``I'm so nervous for the next round I'm not going to be using it (the FeedbackWriter condition). And I'm like, it’s gonna suck.''}

\subsubsection{FeedbackWriter interface facilitated feedback provision}
Participants appreciated FeedbackWriter for having an integrated space to provide feedback. They like being able to provide in-text comments compared with writing on a separate page, which reduced context switching during grading (P1, P9) and made it easier to recapitulate their reasoning when a regrade request arrived (P9). Participants especially liked having the rubric-aligned feedback boxes available as a checklist to ensure comprehensive evaluation. As P11 said, \textit{``Because all of them (rubrics) are there, I'm less likely to miss ... it's not like I have to split the screens (as in previous semesters). That's kind of glitchy.''} When grading final drafts, participants valued seeing the students' edits alongside earlier feedback to ensure continuity from first to final draft and to align comments with students' improvements (P1, P4, P5).

\subsubsection{AI suggestions help TAs provide feedback more effectively and thoroughly.} 
Despite occasional mistakes, all participants reported that the AI suggestions were accurate enough to be helpful and time-saving. As P2 noted, \textit{``the quality of feedback would be higher \dots and much quicker with the AI.''} And P11 said, \textit{``I was definitely more consistent with the AI.''} Four recurrent benefits emerged across TAs.
First, participants appreciate that AI suggestions caught student mistakes that they overlooked on their own.
In contrast, several participants from the baseline condition reported that when they graded the final draft, they realized having missed student mistakes on the first draft (P2, P3, P8). 

Second, participants reported that checking the AI judgments and highlights helped them double-check and think more thoroughly (P1, P5, P10). 
Especially when participants find the AI judgment different from theirs, they would use the AI highlights to understand the AI reasoning and double check their own judgments. As P5 notes, \textit{``Those few times I don't agree (with the AI judgments) \dots That let me think about it more to see if the student did meet it or not. It let me do a more thorough job.''} They found the surfaced cases help them reconsider the rubric's intent (P1, P11), notice misalignment in their interpretations (P4), and flag items for the instructor's clarification (P5). This helped TAs \textit{``grade exactly to the rubric''} (P1), improving rubric understanding and cross-grader consistency.

Third, participants appreciated AI feedback's Socratic scaffolding without telling, while their own feedback always \highlight{gave} away the answers (P2, P9, P10), and made it easier to include brief positive reinforcement they might otherwise omit (P2). \highlight{They also appreciate the positive feedback suggestions for reinforcing what the student did well (P1, P3).} Although participants are aware of the benefits of these strategies, they usually don't have time to include the additional considerations (P1, P2, P5, P9).
For example, P1 said, \textit{``Sometimes I would like to do that, but … it takes time to write the positive feedback … so that is also what was nice about the AI feedback, because you could just put in a few positive comments, and it wouldn't take like any more time.’’}

\subsubsection{AI mistakes had limited impact due to TA agency and predictable error patterns}
Participants described a verification-first stance, where they made their own judgment before consulting AI suggestions as a second opinion. They appreciate that the AI suggestions are easy to modify (P1, P2, P5), both AI feedback and historic feedback is available as options (P2, P10, P11), with none is adopted by default (P3, P8). 
They also recognized a consistent error pattern: AI tended to be stricter. TAs preferred this conservative trend to an over-lenient alternative (P1, P3, P5, P10), since it's easier for them to verify false positives than to identify missed issues. As P1 said, \textit{``I’d prefer them to be more strict... If it said it was good when it wasn’t, I might skip over something.''}
When revisions were needed, TAs said the main reason is usually the worry of AI suggestions containing too few hints. However, they generally preferred this indirectness. TAs find it easy to add actionable details themselves, but much harder to craft helpful feedback that doesn’t reveal the solution, both when editing AI feedback and when writing comments themselves (P1, P2, P5, P9, P10).

\section{Discussion}
\subsection{What Makes FeedbackWriter Effective?}
Our findings suggest that when it's easy for TAs to review, adopt and edit AI feedback suggestions, AI-mediated feedback has higher benefits compared to human-only feedback. Specifically, students who received AI-mediated feedback had higher quality revisions. AI-mediated feedback also demonstrated more desirable properties of effective feedback, such as actionability and promotion of independent learning. 

We attribute FeedbackWriter's benefits to three reinforcing factors in design:
\begin{enumerate}
    \item  \textbf{AI complements human evaluators' expertise} in being comprehensive about all rubric items and crafting feedback messages that align with desirable feedback properties.
    \item \textbf{The interaction design of FeedbackWriter} makes it easy for human evaluators to review AI suggestions. In particular, FeedbackWriter aligns with human evaluators' cognitive processes of reading essays and providing feedback, and \textbf{provides affordances at each step} for the human evaluator to correct AI mistakes and adopt AI suggestions. 
    \item \textbf{We iteratively improve assignment rubrics with the instructors.} The detailed rubrics make it possible both for AI to make accurate judgment and generate fine-grained feedback messages, and for the human evaluators to parse and use AI suggestions more efficiently. 
\end{enumerate}

We further elaborate on each of the factors---\textbf{AI complements human evaluators' expertise:}
For lengthy essay assignments with extensive rubrics, human evaluators can easily overlook criteria. 
By anchoring each comment to a specific rubric item, FeedbackWriter promotes complete and consistent coverage. 
TAs shared that AI was often stricter than they were, catching errors they would have overlooked. This was particularly helpful for first-draft feedback, since students would have received more feedback to improve their work. Moreover, TAs considered the AI feedback to be better worded. 
Sometimes, even after TAs judged whether each rubric was met, it took time and effort to craft the feedback message. The AI’s drafts provided strong starting points and rhetorical templates, accelerating comment writing and helping TAs frame guidance for students.
\highlight{Our findings suggest that this support is necessary. Even when TAs understood effective feedback strategies, time constraints limited consistent application. Several participants who used FeedbackWriter in Assignment 1 wanted to adopt AI’s strategies when writing their own feedback, yet could not do so regularly. Under time pressure, TAs prioritized error correction over other high-quality properties, such as promoting independent learning. This helps explain why, after seeing AI suggestions in Assignment 1, participants in the baseline condition for Assignment 2 still created feedback of lower quality than in the AI-mediated condition.
In contrast, human-AI collaboration offloads tedious searching and phrasing, which frees TAs to focus their limited time and cognitive effort on higher-level instructional interactions.
}

\textbf{The interaction design of FeedbackWriter is critical for effective AI-mediated feedback:} 
FeedbackWriter is deliberately aligned with human evaluators' natural workflow. 
Prior work shows that AI-generated content can be hard to parse \cite{lu2023readingquizmaker, park2024promise, woodruff2024knowledge}. 
Presenting a single, integrated feedback block for the entire essay would be difficult for TAs to edit. TAs would need to read the essays and the long feedback block, while figuring out the evidence for each comment, raising cognitive load.
FeedbackWriter decomposes feedback at the rubric level and highlights the specific essay sentences that informed each AI judgment. This rubric-aligned, evidence-revealing design makes AI suggestions easier to audit, correct and adopt, which provides additional information to human evaluators without overburdening them. 

\textbf{Detailed rubrics iteratively refined by instructors:} Between Assignment 1 and 2, we found that AI's judgment was more accurate for Assignment 2, resulting in significantly fewer overrides by the TAs. 
We attribute this to deliberate rubric refinement with instructors, e.g., making criteria explicit and enumerating the ideas students must demonstrate to count as "meeting" the criteria.
This process is both helpful for the AI to make judgments at scale, but also helpful to increase the consistency among the TAs. We recommend this as a good practice to externalize instructors' tacit knowledge into precise rubric language. 
For example, the AI-human disagreements can be used to flag under-specified rubric items for targeted revision. \highlight{This echoes findings from prior work that improved rubric items could improve grading and feedback quality \cite{yuan2016almost}.}

\subsection{What about AI-Only Feedback?}
This work compares AI-mediated feedback with human-only feedback. A natural alternative is AI-only feedback, which has the obvious advantage of saving instructors' time. In our study, TAs spent similar amount of time in both FeedbackWriter and the Baseline conditions. The user interviews offered insights to further improve usability of the system to save time, e.g., grouping rubrics. However, the current system didn't result in a time reduction, which makes AI-only more appealing if time savings is a priority. 

Future work should directly compare AI-mediated feedback with AI-only feedback. \textbf{This study provides suggestive evidence on why AI-mediated feedback might outperform AI-only feedback.} First, we found that for 88.7\% of the time, AI's judgment aligns with the TAs' judgment (TAs retained the AI's judgment), but 11.3\% of the time, TAs overrode AI judgment, suggesting nontrivial error risk of using an AI-only pipeline. Second, beyond AI-generated feedback, TAs also adopted historic feedback 1,348 times throughout the study, where they regarded the historic feedback to be trustworthy, and sometimes more conceptually beneficial. Moreover, TAs made deliberate edits to adopted AI or historic feedback. There was a total of 806 feedback comments that were further modified after adoption. Together, these patterns suggest that AI-mediated workflows can couple scale with expert oversight, yielding higher reliability and more targeted guidance than AI-only. 
Additionally, using AI-only feedback may bring ethical concerns and negatively impact students' motivations and trust. 

We also suggest a future direction using TAs' feedback edit history to fine-tune and improve AI performance. Future work could also explore real-time adaptation\highlight{---}where the system incorporates TA edits on the fly to improve AI's performance. 


\subsection{Suggestions on Rubric Development}

Clear, operational rubrics are foundational to reliable AI judgment and, consequently, to effective AI-mediated feedback. Decades of research on automation in HCI \cite{horvitz1999principles, sarter1995world, endsley1995out}, along with more recent work in human-AI interaction \cite{kocielnik2019will, wang2023watch, poursabzi2021manipulating}, document the ``out-of-the-loop'' problem: automation can dminish users' situation awareness; when it fails, humans pay a steep cost to re-enter the loop; and ambiguous system states increase confusion and recovery effort. Because providing feedback is itself cognitively demanding, well-specified rubrics help reduce extraneous cognitive load and enable evaluators to more efficiently audit, calibrate, and selectively adopt AI-generated suggestions. 
Moreover, the proces of refining rubrics parallels knowledge engineering efforts in cognitive tutor development, where extracting and formalizing cognitive models is central to building effective instructional systems \cite{lovett1998cognitive, koedinger1997intelligent, aleven2006cognitive}

We recommend a data-informed, continuously updated knowledge engineering and rubric refinement process:
\begin{itemize}[nosep,leftmargin=*]
\item \textbf{Iterate from real data.} Expert blind spots \cite{wang2021seeing} can make criteria hard to fully articulate. Using AI as a first-pass evaluator can surface systematic disagreements between AI and instructors that reveal where rubrics lack precision.
\item \textbf{State canonical criteria, acceptable variants, and use examples.} For example, a rubric may require students to understand what is "substitutes in consumption". If a student writes only that two goods are "substitutes", AI often marks this correct, while instructors will not. 
As data accumulates, rubrics can also be supplemented with examples to further specify the criteria. 
\item \textbf{Encode strictness by rubric type and consider dynamic rubrics.} Instructors have expressed different "strictness" levels they would like to impose on different rubrics. For example, for rubrics that are about definitions, instructors would like to see specific terminology used. On the other hand, for rubrics that are about the analysis of a problem, the solution space can be much bigger. Future work could also explore "dynamic" rubrics, which align with instructors' expectations of students' mastery levels on different types of knowledge \cite{koedinger2012knowledge}.  
\end{itemize}

\subsection{Suggestions on Human-AI Systems for Feedback Provision}\label{sec:designsuggestion}
Feedback provision is a creative and cognitively demanding task. \highlight{Our results show that TAs perceive the AI assistance to be helpful and preferable. With comparable grading time in both conditions, TAs in the FeedbackWriter condition constructed more comprehensive and higher-quality feedback, indicating more efficient feedback provision.} From the study, \highlight{One main source of time spent is on checking the AI suggestions one by one, and making manual edits. Here we discuss how} human-AI interaction design can be further improved to support feedback provision:
\begin{enumerate}
    \item \textbf{AI can be stricter, it is easy for TAs to correct overly strict AI.} When AI tagged a rubric item as missing, TAs found it easy for them to correct the AI's mistake. On the other hand, when AI tagged a rubric as satisfied, it was harder for TAs to realize it. This suggests that if there is a trade-off between false positives and false negatives, AI judgment can be made stricter. 
    \item \textbf{Writing guiding questions and implicit hints is more difficult for TAs.} TAs found that after they made the judgment on whether a rubric is satisfied, crafting a hint message that guides the student but doesn't reveal the correct answer is difficult. They often find inspiration in AI-generated messages. This suggests that AI feedback can be fine-tuned to be more implicit to be complementary to TAs' abilities.
    \item \textbf{More intra-evaluator personalization can be done on the fly.} We found that TAs had individual styles of writing feedback comments. For example, some TAs often began with "I would have liked to see you discuss ...". The system can potentially update AI-generated feedback messages to align with the TAs' personal preferences. 
    \item \textbf{Displaying confidence levels on rubrics may support skimming.} We found that the human-AI agreement rate differs by rubrics. For some rubrics, AI judgments are almost 100\% retained, whereas for others, they are more often overridden by TAs. Displaying the agreement rate might \highlight{help TAs allocate their time more efficiently}. 
\end{enumerate}

\subsection{\highlight{The Implementation Scope of FeedbackWriter}}

In our study context, the writing assignment is best understood as a Writing-to-Learn \xinyi{(WTL) prompt, as characterized in prior work \cite{finkenstaedt2023portrait, klein1999reopening, reynolds2012writing}. WTL prompts use writing to evaluate and improve }students’ conceptual understanding \xinyi{\cite{reynolds2012writing, emig1977writing}}, such as their analysis of an economic phenomenon, while the rhetorical writing quality, such as the strength and coherence of their arguments, is not assessed directly. Although the WTL assignments can be open-ended by, for example, having students find and identify an economic phenomenon in a news article, the writing is expected to follow a pre-defined analytical structure, requiring what needs to go into this analysis. FeedbackWriter generates feedback aligned to that structure through explicit, rubric-based expectations. 

\xinyi{Our results suggest that when success criteria are explicit and operationalized in rubrics, AI-mediated support can increase the coverage and actionability of formative comments, improving revision quality. This aligns with education research showing that rubrics can support both feedback quality and revision when they externalize expectations \cite{hattie2007power, panadero2013use}. 
Systems like FeedbackWriter are} well-suited to writing tasks where the expectations can be articulated. For example, in a statistics course, when students interpret results, or in an engineering course, when students justify design choices against explicit requirements, and may be most effective in foundational gateway courses, where the core concepts to be acquired are more often clearly defined. 

\xinyi{FeedbackWriter may be less effective for writing tasks in which quality is not easily reducible to a shared analytic frame. This includes tasks involving diverse, ill-structured problems \cite{jonassen1997instructional, spiro2019cognitive}, dialogic or reflexive forms of argumentation \cite{bereiter2013psychology, hayes2012modeling}, and rhetorical writing that emphasizes stylistic fluency or narrative flow \cite{reynolds2012writing, flower1979writer}. In such settings\cite{miller1984genre, bazerman1988shaping}, systems like FeedbackWriter may require different representations of quality, such as richer models of argumentation, discourse moves, or dialogic feedback processes). At the same time, teacher-AI collaborative workflows may be especially critical in these contexts, as instructors remain responsible for evaluating dimensions of writing that are difficulty to capture in explicit quality representations.}

Another implication concerns the diversification of assessment formats that systems like FeedbackWriter could afford. Extensive prior work highlights test anxiety in college gateway courses, such as introduction to biology \cite{england2017student}, physics \cite{malespina2022gender}, economics \cite{macri2025effect}, and statistics \cite{wu2022reducing, butakor2024relationship}. By improving the scalability and consistency of feedback on Writing-to-Learn assignments, technologies like FeedbackWriter may support assessment modalities that allow students to demonstrate learning at their own pace, without imposing prohibitive workload costs on instructors. In this way, systems like FeedbackWriter could help broaden the range of valid assessment practices in large-enrollment courses.

\subsection{Limitations}
(1) In this work, we used an AI-based evaluator to evaluate essay quality. We discussed the rationale, e.g., human raters have higher variance and the AI pipeline achieves high accuracy in comparison to human expert judgements. Future work should explore obtaining reliable human-evaluation scores and examine whether the findings still hold. 
(2) We compared AI-mediated feedback with human-only feedback. There is a lack of comparison with AI-only feedback. We offered hypotheses on the performance of AI-only feedback. Given the time savings factor, future work should compare AI-mediated feedback with AI-only feedback. 
(3) \xinyi{In this work, we focused on understanding how AI-mediated feedback influences students’ revision quality. Because each revision provides only a single learning opportunity for a given knowledge component, the expected effect size on downstream learning outcomes is likely smaller than the effect observed for revision quality \cite{lipsey2012translating}. Detecting such learning gains would therefore require higher instructional dosage, such as repeated revision practice across multiple assignments, as well as larger sample sizes.
Future work should investigate how AI-mediated feedback supports student learning when embedded within sustained practice over time and evaluated at scale.}
(4) \highlight{We focused primarily on TAs' perspectives on AI-mediated feedback.} The main student-facing data collected in this study were performance measures from assignments and quizzes; we did not survey or interview students about their experiences receiving AI-mediated feedback \highlight{or their attitudes towards instructors' use of AI. Future work should examine both students' and instructors' perceptions of AI-assisted feedback, including its effects on social dynamics such as accountability, trust, and TA-student interaction. In addition, although students in our study were aware that TAs used an AI-powered tool, they were not informed which specific feedback comments were AI-suggested. In practice, TAs could adopt an AI suggestion while still substantially revising or reframing the feedback. Future research should investigate how awareness of AI involvement at the level of individual feedback comments influences students’ engagement with feedback and their motivation to revise.}




\section{Conclusion}
We introduce FeedbackWriter, a system that provides AI suggestions to help human evaluators provide feedback to knowledge-intensive essays. We also report findings from a randomized trial involving 354 students and 11 teaching assistants (TAs).  
Our findings suggest that when making it easy for TAs to review, adopt and edit AI feedback suggestions, AI-mediated feedback has higher benefits compared to human-only feedback. Specifically, students who received AI-mediated feedback had higher quality revisions. AI-mediated feedback also demonstrated more desirable properties of effective feedback, such as actionability and promotion of independent learning. 
We attribute FeedbackWriter's benefits to three reinforcing factors: 1) AI complements human evaluators' expertise in being comprehensive about all rubric items and crafting feedback messages that align with desirable feedback properties; 2) The interaction design of FeedbackWriter makes it easy for human evaluators to review AI suggestions; 3) The detailed rubrics iteratively refined by the instructors make it possible both for AI to generate accurate judgment and fine-grained feedback, and for the human evaluators to parse and use AI suggestions more efficiently. 
\begin{acks}
This work is funded by NSF Grants IIS-2442990 and IIS-2302564. 
We would like to acknowledge Aditya Mahesh and Zejia Shen for contributing to the early development of the system; Aditya Mahesh and Jiarui Zhang for grading post-test 1 and 2 respectively; Lu Wang, Anne Gere, Inderjeet Nair, and members of the Michigan Lifelong Learning Lab for feedback on the project. In addition, we sincerely thank all the instructors and TAs who participated in our study for sharing their thoughtful feedback on our system, and the reviewers for their valuable comments. 
\end{acks}

\bibliographystyle{ACM-Reference-Format}
\bibliography{ref}

\clearpage
\appendix
\section{Feedback Analysis}

\subsection{Feedback Categorization}
\label{appendix:categorization}

Following the work of Patchan et al. \cite{Patchan_Schunn_Correnti_2016}, we categorized feedback into 4 categories: summary, praise, problem, and solution.

\subsubsection{Summary} Feedback of this category should be a comment that describes or summarizes the student’s argument or written content in the essay. A typical summary starts with phrases such as "you've mentioned/discussed.”

Examples:
\begin{itemize}
    \item "You mentioned government expenditure..."
    \item “Your explanation of market failure highlights inefficient resource allocation.”
\end{itemize}

\subsubsection{Praise} Feedback of this category should describe a positive feature of the student’s essay.

Examples:
\begin{itemize}
    \item "Great job identifying the shifts and their resulting changes in equilibrium price and quantity."
    \item “Your essay is well-written and free of grammatical errors, which helps maintain clarity and professionalism in your writing.”
\end{itemize}

\subsubsection{Problem} Feedback of this category should describe what is wrong with the student’s essay. We considered both statements and questions that point out an issue as problem-type feedback.

Examples:
\begin{itemize}
    \item "...instead of saying that there is increased production as this isn't quite accurate."
    \item "Did you present any information in your essay that was drawn from the article?"
    \item “Market Failure occurs, when absent governmental intervention, the market inefficiently fails to maximize society's total surplus.” \textit{(Note: This feedback is a problem, because it states a correct definition to contrast with the student’s answer, which indirectly points out an issue.)}
\end{itemize}

\subsubsection{Solution} Feedback of this category should describe actions to fix a problem or improve the essay quality. Comments with the verb *ould—should, could, would—and phrases such as “be sure to” or “remember to” are typical solutions. We also considered guiding questions that hint at potential actions as solutions.

Examples:
\begin{itemize}
    \item “I'd like you to explain how exactly AI is nonrival and excludable.”
    \item "Remember to discuss what implications this policy has for the welfare of the producer in the almond milk market."
    \item “Compare your market failure to another market failure from a type of good. How does the efficient quantity differ from the market quantity with these two types of goods?”
\end{itemize}

\subsection{Four-Metric Rubric for Feedback Quality Evaluation}
\label{appendix:feedback_rubric}
We designed a rubric that measures feedback quality in 4 dimensions: content accuracy, promotion of independent learning, actionability, and tone and supportiveness. Each metric is binary. If the feedback meets the description, it receives 1 point. If it doesn't, it receives 0 point.


\begin{table*}
\centering
\caption{Rubric for feedback quality evaluation based on 4 metrics: content accuracy, promotion of independent learning, actionability, and tone and supportiveness.}
\footnotesize
\setlength{\tabcolsep}{4pt}
\renewcommand{\arraystretch}{1.15}

\begin{tabular}{|p{2.1cm}|p{4.1cm}|p{4.1cm}|p{4.1cm}|}
\hline
\textbf{Metric} & \textbf{Description} & \textbf{Positive Examples} & \textbf{Negative Examples} \\
\hline

Content Accuracy
&
This metric measures whether the graders show strong subject-matter knowledge.
\newline\newline
\textbf{Met:} The grader gives feedback that correctly reflects economic principles/concepts based on the rubric. The grader demonstrates strong subject-matter knowledge.
\newline\newline
\textbf{Not Met:} The grader provides feedback that mostly or completely misrepresents economic principles/concepts based on the rubric. The grader lacks subject-matter knowledge.
&
``Quantity supplied and quantity demanded in the almond milk market will increase along their respective curves, not supply and demand.''
\newline\newline
``Remember to identify the consumers and producers of each market. Evaluate who drives the demand in the water market, address both consumers.''
&
``You say that `the demand curve for oat and soy milk will shift right', but these goods are substitutes, so they should shift together.''
\newline
\textit{(Note: The grader misidentifies the concept of substitutes. The demand for one good depends on the price of the other. For example, if the price of soy milk rises and its quantity demanded falls, the demand for oat milk should increase, shifting its curve to the right.)}
\\
\hline

Promotion of Independent Learning
&
This metric measures whether the graders design feedback that elicits a student's independent thinking.
\newline\newline
\textbf{Met:} The grader gives a suggestion without revealing answers or concepts through indirect hints, suggestions, or guiding questions. Feedback is designed to help the student recognize and address issues independently.
\newline\newline
\textbf{Not Met:} The grader directly reveals answers, concepts, or mistakes without indirect suggestions, hints, or guiding questions to encourage reflection.
&
``Remember to discuss what external marginal cost is and how that relates to social marginal cost.''
\newline\newline
``Can you explore how the changes in demand for different types of milk might collectively impact water demand before the policy?''
\newline\newline
``Make sure you understand the difference between demand and quantity demanded when discussing adjustments in the market.''
&
``There will not be a shortage in the alternatives market, the market will just adjust to equilibrium.''
\newline\newline
``There are no comparisons to similar failures within the essay.''
\newline
\textit{(Note: The grader only states the mistake without indirect suggestions to promote independent thinking.)}
\newline\newline
``You should explicitly mention that almond milk requires a greater amount of water as an input than the alternatives.''
\\
\hline

Actionability
&
This metric measures whether the feedback is clear and specific.
\newline\newline
\textbf{Met:} The grader consistently gives feedback that is specific and personalized. The feedback includes concrete questions, examples, explanations, or hints. The language is clear for students to understand next steps.
\newline\newline
\textbf{Not Met:} The feedback contains vague or unclear examples, explanations, or hints. The feedback is not personalized based on the student's writing. The language lacks clarity for students to understand next steps.
&
``Remember to discuss what external marginal cost is and how that relates to social marginal cost.''
\newline\newline
``Quantity supplied and quantity demanded in the almond milk market will increase along their respective curves, not supply and demand.''
\newline\newline
``Can you explore how the changes in demand for different types of milk might collectively impact water demand before the policy?''
&
``You should list the possible solutions.''
\newline\newline
``How do you calculate the tax revenue the government generates?''
\newline
\textit{(Note: This feedback asks an unclear guiding question, so students may not understand what ``calculate the tax revenue'' means.)}
\newline\newline
``Be sure to explain the general idea of both positive and negative externalities.''
\\
\hline

Tone and Supportiveness
&
This feedback measures whether the graders’ tone is appropriate.
\newline\newline
\textbf{Met:} The grader communicates feedback in a suggestive and respectful tone. The feedback has a good balance between acknowledging attempts, summarizing, explaining, and providing suggestions for improvement.
\newline\newline
\textbf{Not Met:} The feedback doesn’t contain any suggestions, explanations, and acknowledgement, or uses a condescending tone (e.g., too many negative comments). The feedback may discourage student’s motivation and confidence in learning economics.
&
``Make sure to properly define market failure and include ALL parts of the definition. I recommend going back to lecture slides for this.''
\newline\newline
``I appreciate your discussion of social marginal cost/benefit, but I encourage you to compare this to the private marginal cost/benefit instead of just mentioning that it `is changed'.''
&
``Common resources lead to overconsumption, not overproduction.''
\newline
\textit{(Note: The grader only restates a definition without giving any suggestions or acknowledgement, which makes their tone cold.)}
\newline\newline
``I noticed there weren't any in-text citations in your essay, but there was a quote. The quote is not at all relevant though.''
\\
\hline

\end{tabular}
\end{table*}

\section{Interview Guide}
In this session, we’d like to understand how you use the system to review student writing and provide feedback. 

[Warmup questions]
\begin{itemize}
    \item In general, how was your grading experience?
    \item Can you briefly describe your general process for providing feedback using the system?
\end{itemize}

[Think aloud]
Please imagine this is your first time grading this assignment. Here is the first student writing we will be looking at. It would be helpful if you could share your screen, so that we could better understand what you are referring to. In your process, it’d be really great if you could think out loud on what you notice, why you’re choosing certain actions, and how you interpret the system’s suggestions. There are no right or wrong answers, but we are interested in your reasoning.

Question/observations:
\begin{itemize}
    \item How did you combine reading, evaluating, feedback provision and checking AI suggestions in the process?  (e.g. They might read through the essay before evaluation or give feedback as they read)
    \item I noticed that you used AI feedback and historic feedback interchangeably. How did you decide which to use? 
    \item For [the specific feedback], could you compare the historic feedback with the AI feedback? 
    \item (If not satisfied with the feedback suggestions) Could you elaborate on how you manually made the edits on this feedback?
    \item (If they modified AI suggestions) Can you explain what made you disagree with the AI’s judgment?
    \item How did you use the highlights on the essay?
    \item Did you ever adjust the highlights/judgment/feedback? When would/wouldn’t you adjust?
    \item When you see the AI suggestions (highlights, judgments, feedback) to be incorrect, how would you deal with them?
    \item If you have toggled a judgment, did you find the feedback anchoring at the correct place?
    \item How does grading with this system compare to your experience using Canvas or other tools in previous semesters?
    \item What are some challenges or frustrations you experience during the process, if any?
    \item If you could change one thing to make your workflow smoother, what would it be?
\end{itemize}

[Conclusion]
Thank you so much for participating in this study. Before we end, are there things that you wanted to share or discuss but we haven’t touched upon?


\end{document}